\documentclass[%
 reprint,
 amsmath,amssymb,
 aps,
]{revtex4-2}

\usepackage{graphicx}
\usepackage{dcolumn}
\usepackage{bm}
\usepackage{hyperref}
\usepackage{dsfont}
\usepackage{subfigure}
\usepackage{comment}
\usepackage{color}

\def\msun{\,{\rm M}_\odot}
\def\zsun{\,{\rm Z}_\odot}
\def\rsun{\,{\rm R}_\odot}
\def\dd{\,{\rm d}}
\def\d{{\rm d}}

\begin{document}

\preprint{APS/123-QED}

\title{Flipping spins in mass transferring binaries\\
and origin of spin-orbit misalignment in binary black holes}

\author{Jakob Stegmann}
 \email{StegmannJ@cardiff.ac.uk}
\author{Fabio Antonini}%
 \email{AntoniniF@cardiff.ac.uk}
\affiliation{%
 Gravity Exploration Institute, School of Physics and Astronomy,\\Cardiff University, Cardiff, CF24 3AA, UK
}%

\date{\today}

\begin{abstract}
Close stellar binaries are prone to undergo a phase of stable mass transfer in which a star loses mass to its companion. Assuming that the donor star loses mass along the instantaneous interstellar axis, we derive the orbit-averaged equations of motion describing
the evolution of the donor rotational angular momentum vector (spin) which accompanies the transfer of mass. We consider: (i) a model in which the mass transfer rate is constant within each orbit and (ii) a phase-dependent rate in which all mass per orbit is lost at periapsis. In both cases, we find that the ejection of $\gtrsim 30$ per cent of the donor's initial mass causes 
its spin to nearly flip onto the orbital plane of the binary, independently of the initial spin-orbit alignment. Moreover, we show that the spin flip due to mass transfer can easily dominate over tidal synchronisation in any giant stars and main-sequence stars with masses $\sim1.5$ to $5\msun$. Finally, the general equations of motion, including tides, are used to evolve a realistic population of massive binary stars leading to the formation of binary black holes. Assuming that the stellar core and envelope are fully coupled, the resulting tilt of the first-born black hole  reduces its spin projection onto the orbit normal by a factor $\sim\mathcal{O}(0.1)$. This result supports previous studies in favour of an insignificant contribution to the effective spin projection, $\chi_\text{eff}$, in binary black holes formed from the evolution of field binaries.
\end{abstract}

\maketitle


\section{Introduction}\label{sec:introduction}
A large number of stars are found to be  in close binary systems. The fraction of main-sequence stars which are bound to one or more companions ranges from $\sim40$ per cent in the case of solar-type stars up to $\sim90$ per cent for massive O-type stars \citep{2017ApJS..230...15M,2012Sci...337..444S,1991A&A...248..485D}. A substantial fraction of binaries move on close orbits with orbital periods less than $\sim10^3$ to $10^4\,\text{days}$. Compared to a life in isolation, their evolutionary pathways are significantly altered as they can undergo a phase of mass transfer in which they exchange a large amount of mass and rotational angular momentum with their close companions \citep{1971ARA&A...9..183P}. Mass transfer between binary members is responsible for a set of astrophysical phenomena such as X-ray binaries \citep{1973A&A....24..337S} and millisecond pulsars \citep{1991PhR...203....1B}. Moreover, the larger and more evolved but paradoxically less massive members of Algol-type eclipsing binaries are thought to become so during a phase of mass transfer to their companions \citep{1955ApJ...121...71C}. 

A number of  binary stellar evolution codes exist that allow to simulate the life of binary stars including mass transfer phases along with other binary effects such as mass accretion, common-envelope evolution, supernova kicks, and angular momentum losses (e.g., {\fontfamily{qcr}\selectfont BSE} \citep{2002MNRAS.329..897H}, {\fontfamily{qcr}\selectfont StarTrack} \citep{2008ApJS..174..223B}, {\fontfamily{qcr}\selectfont MESA} \citep{2011ApJS..192....3P}, {\fontfamily{qcr}\selectfont binary\_c} \citep{2004MNRAS.350..407I}).
Regarding the mass transfer, these codes typically build upon two assumptions. Firstly, the effect of any orbital eccentricity is neglected during the mass transfer phase. For circular orbits, there exists the well-established Roche lobe limit which a star's radius has to exceed so that it loses mass to its companion \citep{1983ApJ...268..368E}. In turn, modelling the mass transfer rate on eccentric orbits in which the orbital separation oscillates is extremely difficult since mass transfer might occur partially within each orbit at varying rate and does back-react on the orbital elements changing the eccentricity and semi-major axis. Recently, several promising attempts have been made to solve these difficulties \citep{2019ApJ...872..119H,2016ApJ...825...70D,2016ApJ...825...71D,2007ApJ...667.1170S}. Secondly, the rotational angular momentum vectors (spins) of the binary stars are assumed to be aligned with the orbital axis. This assumption has partly been made due to simplicity and partly because tidal interactions between the binary members are believed to diminish any spin-orbit misalignment \citep[e.g.,][]{1981A&A....99..126H}. However, there is observational evidence of close binaries with highly inclined spin axes suggesting that tides are not in all cases able to align the spins with the orbital axis \citep[e.g., BANANA survey,][]{2009Natur.461..373A}.

In this paper, we question the second assumption of spin-orbit alignment. Based on the work of \citet{1983ApJ...266..776M}, we will show that if a binary undergoes a phase of mass transfer the mass-donating star actually loses rotational angular momentum in a way that causes its spin vector to flip onto the orbital plane. 

We will apply this result to the evolutionary pathways of {massive} stars in close binaries. Isolated in the Galactic field, these systems have been proposed as the progenitors of a formation channel \citep{2020arXiv200611286K,2016Natur.534..512B,2016MNRAS.458.2634M,2012ApJ...759...52D} leading to the binary black hole (BBH) mergers observed by gravitational wave facilities \citep{2019PhRvX...9c1040A}. In this scenario, mass transfer between the two stars precedes a common-envelope phase in which the orbital separation quickly shrinks to values small enough for the black hole remnants to merge in less than $\sim10\,\text{Gyr}$. 


The orientation and magnitude of the black hole spins constitute an important observable to discriminate among the different binary formation channels
\citep[e.g.,][]{2010CQGra..27k4007M,2013PhRvD..87j4028G,2016ApJ...832L...2R,2020ApJ...899L..17Z}. For instance, the LIGO-Virgo detectors are sensitive to the mass-weighted projection of the black hole spins onto the orbital angular momentum,
\begin{equation}\label{eq:chi-eff}
    \chi_\text{eff}=\frac{M_1\chi_1\cos\theta_1+M_2\chi_2\cos\theta_2}{M_{12}}.
\end{equation}
Here, $M_{1,2}$ are the two black hole masses and $M_{12}$ their sum. The spins are usually expressed in terms of the dimensionless spin parameters $\bm{\chi}_{1,2}$ whilst we will use the canonical rotational angular momenta $\bm{S}_{1,2}$ to describe those of their stellar progenitors. Both vectors are related as $\bm{S}_{1,2}=\bm{\chi}_{1,2}GM_{1,2}^2/c$ with $G$ and $c$ referring to the gravitational constant and speed of light, respectively. The angles $\theta_{1,2}=\cos^{-1}\bm{\hat{S}}_{1,2}\cdot\bm{\hat{h}}$ describe the tilts of the spins with respect to the specific orbital angular momentum $\bm{h}$. A viable formation channel has to be compatible with the $\chi_\text{eff}$-distribution of the BBH mergers measured by LIGO-Virgo which peaks around $\chi_\text{eff}\simeq0$ with a slight tendency towards positive values \citep{2019PhRvX...9c1040A,2020arXiv201014527A}. This suggests that the final black hole spins are either small, anti-aligned with each other, or perpendicular to $\bm{h}$.

By means of a population synthesis we will apply the spin dynamics that we derived to the first stable mass transfer occurring in the isolated binary channel. There, we will also take other binary effects such as tidal interactions into account \citep{2001ApJ...562.1012E} in order to investigate whether flipping spins are a prevalent phenomenon or not. Apart from that, we emphasise that the dynamics can be important for any other binary formation channel that might involve a phase of mass transfer, e.g. the triple channel \citep[][]{2020MNRAS.491..495D}, as well as for mass-exchanging stellar binaries in general.

This paper is organised as follows. In Section~\ref{sec:assumptions}, we will outline our basic assumptions. In Section~\ref{sec:spin-evolution}, we will analytically derive the spin dynamics of the mass-losing star. In Section~\ref{sec:TR}, we will discuss the importance of torques emerging from tidal interactions. In Section~\ref{sec:population}, we will present the results of our population synthesis study. Finally, we will summarise our findings in Section~\ref{sec:summary}.

If not stated differently, the magnitude, unit vector, and time derivative of some vector $\bm{V}$ are written as $V=\left|\bm{V}\right|$, $\bm{\hat{V}}=\bm{V}/V$, and $\bm{\dot{V}}=\d\bm{V}/\d t$, respectively.

\section{Basic Assumptions}\label{sec:assumptions}
\begin{figure*}
    \centering
    \subfigure{\includegraphics[scale=0.4]{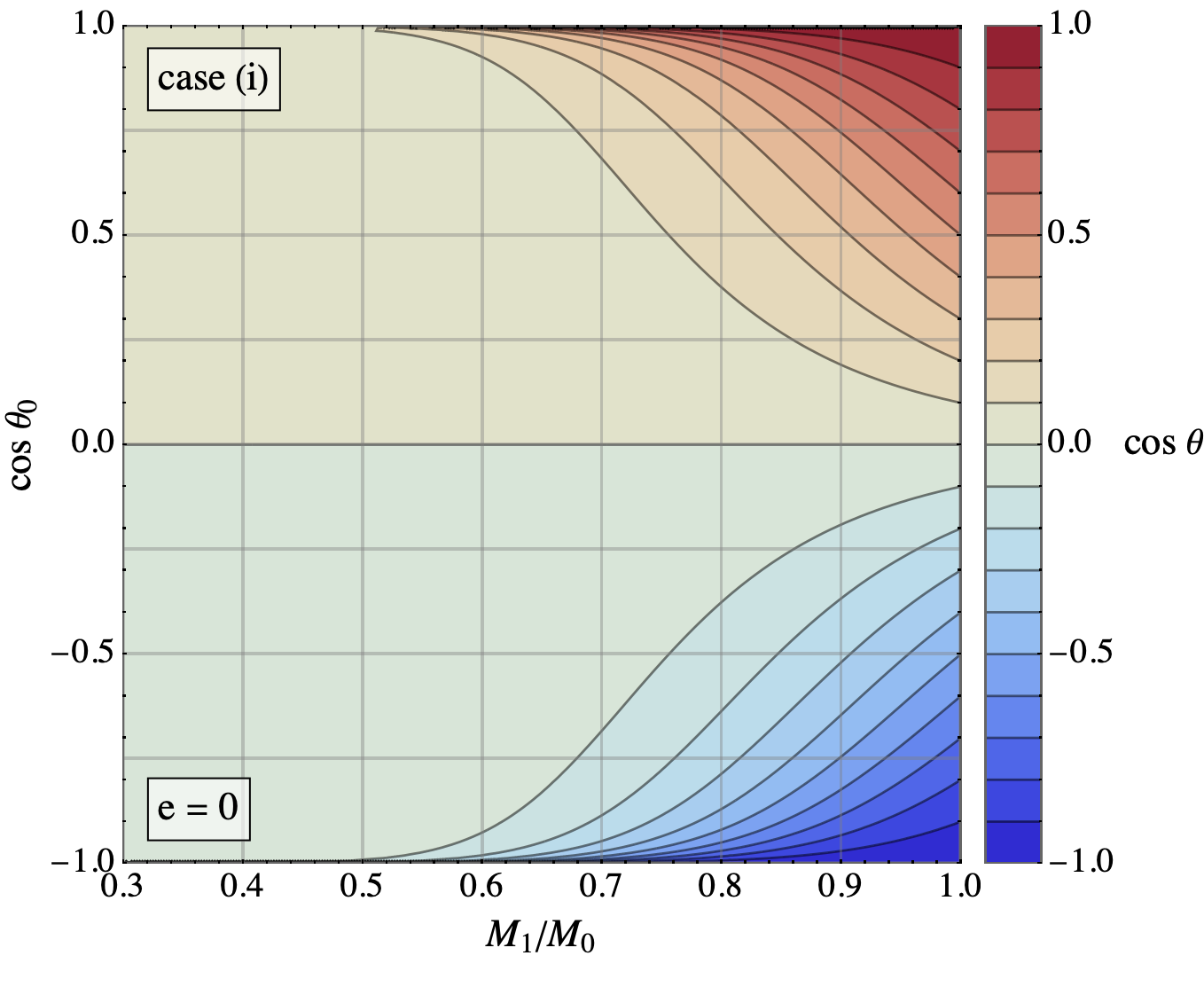}}\quad
    \subfigure{\includegraphics[scale=0.4]{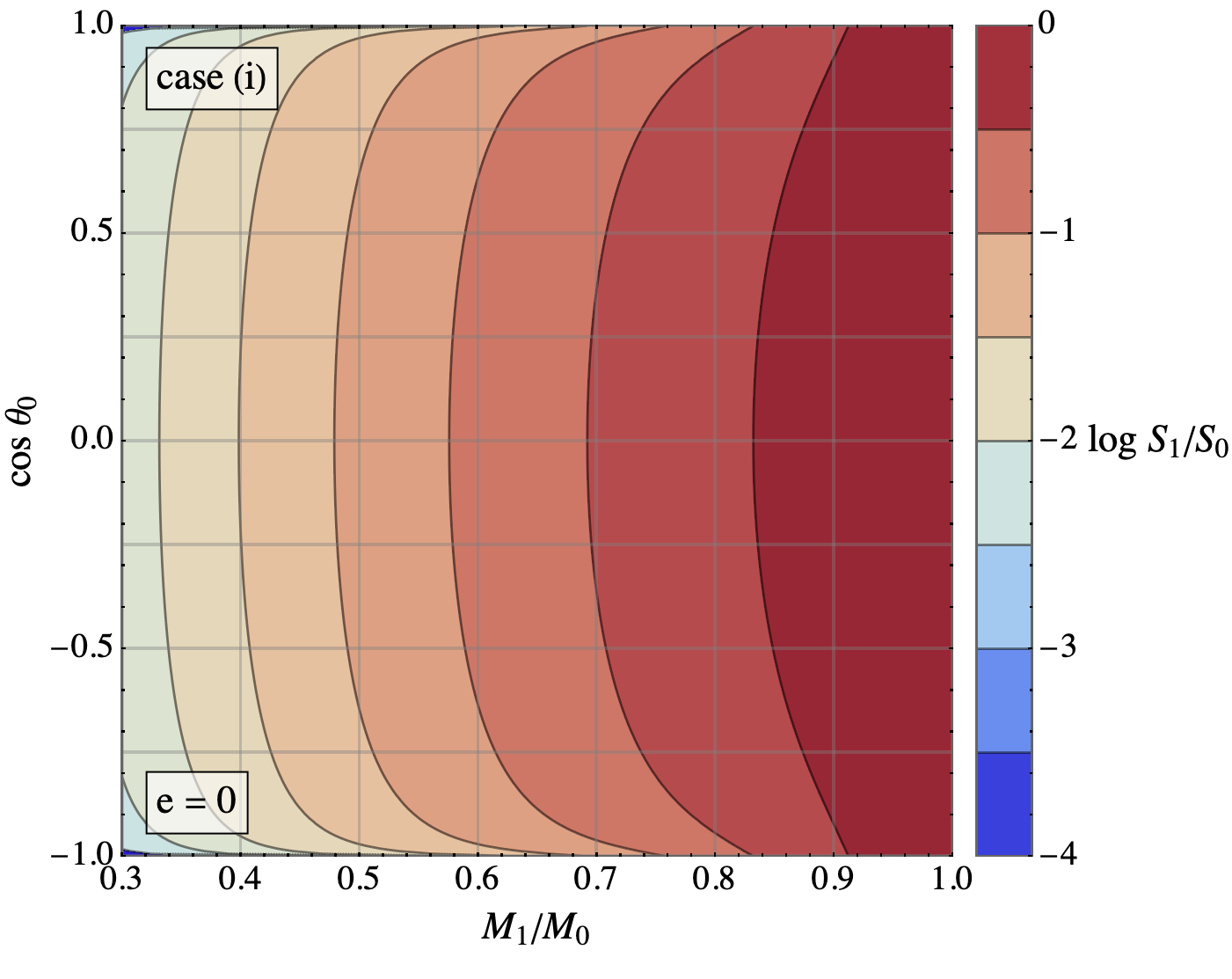}}
    \caption{Donor spin evolution for a constant mass loss rate. We are considering circular orbits ($e=0$). The quantities $M_1$ ($M_0$), $S$ ($S_0$), and $\theta$ ($\theta_0$) describe the current (initial) values of the donor's mass, spin magnitude, and tilt angle with respect to the specific orbital angular momentum $\bm{h}$, respectively. The  {left panel} reveals that as the donor loses mass, i.e. $M_1/M_0$ decreases, any initial donor spin with $\cos\theta_0\in(-1.0,1.0)$ gets flipped towards the orbital plane $\cos\theta_0=0.0$. Meanwhile, the  {right panel} shows how at the same time the spin magnitude efficiently gets damped down.}
    \label{fig:donor-evolution-constant-mass-rate}
\end{figure*}
In this paper, we consider a stellar binary in which one member star transfers mass to the other. We label all quantities related to the mass-losing star (donor) and the mass-gaining star (accretor) with the indices $i=1,2$, respectively. Thus, let $R_i$, $M_i$, $\bm{r}_{i}$, and $\bm{d}=\bm{r}_{2}-\bm{r}_{1}$ denote the stars' radii, masses, the distance between the stellar centers of mass to the binary center of mass, and orbital separation, respectively. Together they carry a specific orbital angular momentum per reduced mass $\mu=M_1M_2/M_{12}$ which is given by
\begin{equation}
    \bm{h}(t)=\bm{d}\times\bm{\dot{d}}\label{eq:L}.
\end{equation}
In terms of the semi-major axis $a$, eccentricity $e$, and total mass $M_{12}$, its magnitude can be written as
\begin{equation}
    h=j\sqrt{GM_{12}a},
\end{equation}
where $j=\sqrt{1-e^2}$. Furthermore, we associate with each star a rotational angular momentum vector (spin) given by
\begin{equation}\label{eq:S}
    \bm{S}_{i}(t)=\sum_k\bm{\rho}{}_{i,k}\times\left(m_k\bm{\dot{\rho}}{}_{i,k}\right)=\sum_k\bm{\rho}{}_{i,k}\times\left(m_k\bm{\omega}{}_{i,k}\times\bm{\rho}{}_{i,k}\right).
\end{equation}
The sums in Eq.~\eqref{eq:S} are taken over all particles with masses $m_k$, absolute positions $\bm{s}{}_{i,k}$, relative positions $\bm{\rho}{}_{i,k}=\bm{s}{}_{i,k}-\bm{r}_{i}$, and angular velocities $\bm{\omega}{}_{i,k}$ that constitute the star $i$ at some time $t$. If we assume for simplicity that the stars retain spherically symmetric shapes during the mass transfer and uniformly rotate at some angular velocities $\bm{\omega}_{i}=\bm{\omega}{}_{i,k}$, one recovers the familiar form
\begin{equation}\label{eq:S-3}
    \bm{S}_{i}(t)=\underline{\bm{\Theta}}{}_{\,i}\cdot\bm{\omega}_{i},
\end{equation}
where $\underline{\bm{\Theta}}{}_{\,i}=\kappa M_i R_i^2\,\underline{\bm{\mathds{1}}}$ is the respective star's total inertia tensor with $\underline{\bm{\mathds{1}}}$ being the identity. Throughout this paper, we set $\kappa=0.08$ for a $n\sim3$ polytrope \citep{1952ApJ...115..562M}.

In general, mass is transferred from the donor to the accretor via their first Lagrangian point $L_1$ once the former fills its Roche lobe \citep{1971ARA&A...9..183P,1975ApJ...198..383L}. That is, the radius of the donor has to expand to the limit approximately given by \citep{1983ApJ...268..368E}
\begin{equation}\label{eq:Roche}
    R_L(t)=dF(q),
\end{equation}
where $q= M_1/M_2$ is the stars' mass ratio and the function $F(q)$ is defined as
\begin{equation}
    F(q)=\frac{0.49q^{2/3}}{0.6q^{2/3}+\ln\left({1+q^{1/3}}\right)}.
\end{equation}
Whenever the donor has grown to that size, $R_1=R_L$, we assume that it loses mass at the point $\bm{R}_{1}= R_1\bm{\hat{d}}$ at a rate $\dot{M}_1=-\dot{M}$ where $\dot{M}>0$ which subsequently gets transferred to the accretor. For simplicity, we assume that the mass transfer is conservative, i.e. no mass is lost from the binary during this process.

\section{Donor spin evolution due to mass transfer}\label{sec:spin-evolution}
In this section we study the spin evolution of the donor based on the work of \citet[]{1983ApJ...266..776M} and \citet[]{2010ApJ...724..546S}. For this reason and for better readability, we will henceforth omit the donor's index $i=1$ ($\bm{S}=\bm{S}_{1}$, $\bm{\omega}=\bm{\omega}_{1}$, $\bm{R}=\bm{R}_{1}$, etc.). Accordingly, consider a general donor quantity $G$ which at some time $t$ can be written as $G(t)=\sum_kG_{k}(t)$, i.e. as a sum over all particles labeled with $k$ that constitute the donor at that time [e.g., Eq.~\eqref{eq:S}]. At a later time $t+\Delta t$, $G$ will be given by $G(t+\Delta t)=\sum_{k'}G_{k'}(t+\Delta t)$, where
\begin{equation}
    \sum_{k'}=\sum_{k}-\sum_{l}
\end{equation}
is the sum over all particles that constitute the respective star at the later time $t+\Delta t$. That is, the donor lost the contribution of the particles labelled with $l$ that left  it within the time interval $\Delta t$. Consequently, the time derivative of $G(t)$ can be written as \citep{1983ApJ...266..776M}
\begin{equation}\label{eq:mass transfer-derivative}
    \dot{G}=\sum_k\dot{G}_k(t)-\lim_{\Delta t\rightarrow0}\frac{1}{\Delta t}\sum_l G_l(t+\Delta t).
\end{equation}
The first term on the r.h.s.~of Eq.~\eqref{eq:mass transfer-derivative} describes the change of $G$ if the mass were held constant, whereas the second term reflects the change due to mass transfer. Insertion of Eq.~\eqref{eq:S} into \eqref{eq:mass transfer-derivative} yields $\bm{\dot{S}}(t)=\bm{\varepsilon}(t)-\bm{\zeta}(t)$ where we defined
\begin{align}
    \bm{\varepsilon}(t)&=\sum_k\frac{\d}{\d t}\left\{\bm{\rho}{}_{k}(t)\times\left[m_k\bm{\dot{\rho}}{}_{k}(t)\right]\right\},\label{eq:tau}\\
    \bm{\zeta}(t)&=\lim_{\Delta t\rightarrow0}\frac{1}{\Delta t}\sum_l\bm{\rho}{}_{l}(t+\Delta t)\times\left[m_l\bm{\dot{\rho}}{}_{l}(t+\Delta t)\right]\label{eq:zeta}.
\end{align}
Physically, the first term $\bm{\varepsilon}(t)$ comprises all external torques applied to the donor spin if its total mass were held constant \citep{1969AIAAJ...7..337H}. These external torques can emerge e.g. from the tidal forces of the companion star and will be discussed in Sections~\ref{sec:TR} and \ref{sec:population}. In order to study the effect of mass loss alone we set $\bm{\varepsilon}(t)=0$ in the remainder of this section. In that case, Eqs.~\eqref{eq:S-3} and \eqref{eq:zeta} yield for the time derivative of the donor spin
\begin{align}\label{eq:zeta-1}
    \bm{\dot{S}}(t)&=-\dot{M}\bm{R}\times\left(\bm{\omega}\times\bm{R}\right)\nonumber\\
    &=-\dot{M}\left[R^2\bm{\omega}-\left(\bm{R}\cdot\bm{\omega}\right)\bm{R}\right]\nonumber\\
    &=-\frac{1}{\kappa}\frac{\dot{M}}{M}S\left[\bm{\hat{S}}-\left(\bm{\hat{d}}\cdot\bm{\hat{S}}\right)\bm{\hat{d}}\right],
\end{align}
\begin{figure*}
    \centering
    \subfigure{\includegraphics[scale=0.4]{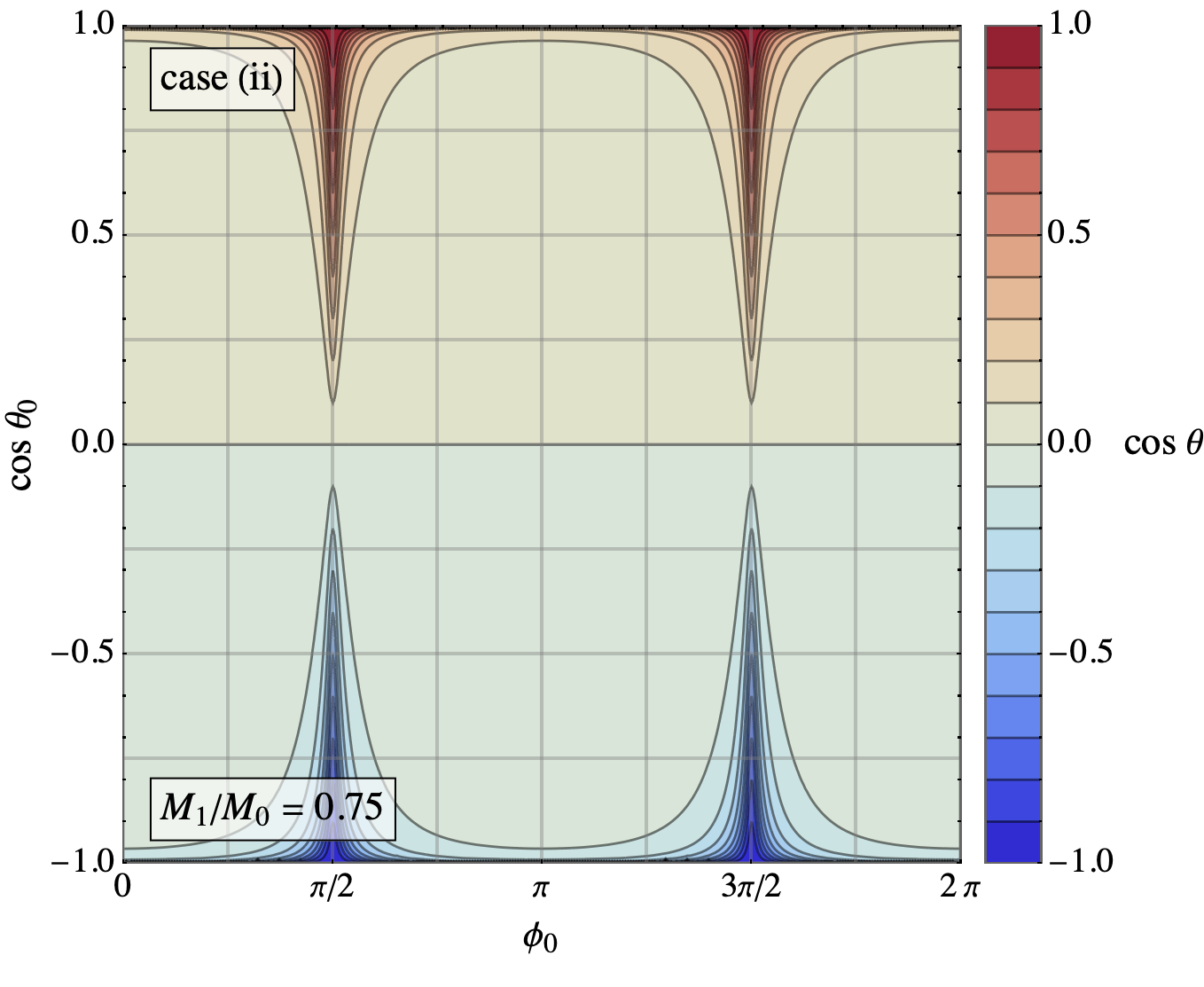}}\quad
    \subfigure{\includegraphics[scale=0.4]{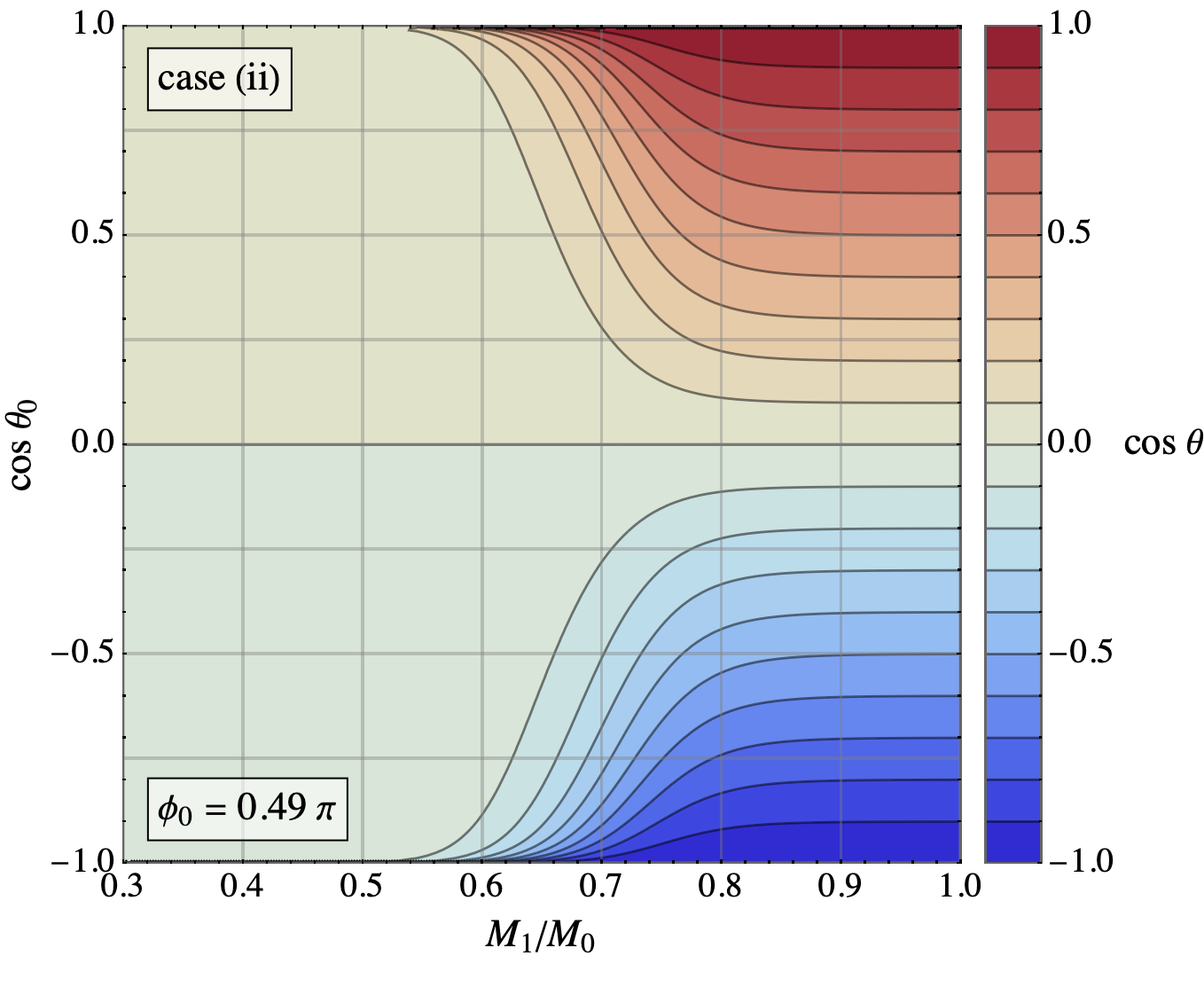}}\quad
    \caption{
    Donor spin evolution given that all mass is lost at the periapsis. The quantities $M_1$, $M_0$, $\theta$, and $\theta_0$ are defined as in Figure~\ref{fig:donor-evolution-constant-mass-rate} whilst $\phi_0$ is the initial azimuthal angle of $\bm{S}$. For a fiducial fractional leftover mass of $M_1/M_0=0.75$, the  {left panel} shows the additional dependency of the tilt angle evolution on the initial azimuthal angle $\phi_0$, revealing that a flip onto the orbital plane is most (least) efficient if it is close to $0$ or $\pi$ ($\pi/2$ or $3\pi/2$). For the latter case ($\phi_0=0.49\pi$), the  {right panel} shows that it requires smaller values of $M_1/M_0$ compared to Figure~\ref{fig:donor-evolution-constant-mass-rate} in order to achieve a significant flip. The opposite would be true if we were to plot the panel for $\phi_0$ around $0$ or $\pi$.}
    \label{fig:donor-flip-delta-mass-rate}
\end{figure*}
since we approximated $\bm{R}=R\bm{\hat{d}}$ to be the first Lagrangian point where the donor loses its particles, i.e. $\bm{R}=\bm{\rho}{}_{l}$. Eq.~\eqref{eq:zeta-1} is equivalent to the derivations of \citet[]{1983ApJ...266..776M} and
partly uses Eq. (23) in \citet{2010ApJ...724..546S}. Here, we make two implicit key assumptions that are commonly used in the literature. Firstly, we assumed that the particles leave the donor through $\bm{R}$ at the donor's rotational velocity, i.e. at a velocity $\bm{\omega}\times\bm{R}$ in the frame co-rotating with the orbital motion \citep{2010ApJ...724..546S}. Secondly, by replacing the angular velocity vector by the spin [c.f. Eq.~\eqref{eq:S-3}] in the last step of Eq.~\eqref{eq:zeta-1} we assumed that the stellar interior transport of angular momentum is efficient enough to align the spins of all parts of the star on a timescale shorter than that of mass loss. Otherwise, the spin direction of some stellar parts, e.g. the core, could in principle decouple from the spin direction of the other parts, e.g. the envelope. 
We discuss the implications of these assumptions in Section~\ref{sec:population}.

The last part of Eq.~\eqref{eq:zeta-1} unveils a clear geometrical interpretation. The first term in the rectangular brackets causes the magnitude $S$ of the donor spin to decrease. Meanwhile, the second term alters the spin direction unless $\bm{\hat{d}}\cdot\bm{\hat{S}}=0$. Thus, depending on the orbital phase and the current spin direction, the second term causes the donor spin vector to either move towards the orbital plane or away from it. In what follows we will  orbit-average the phase-dependent
time-evolution equation~\eqref{eq:zeta-1} in order to investigate which effect dominates over longer timescales. 
 We consider the two cases in which (i) $\dot{M}$ is constant within each orbit and (ii) all mass per orbit is lost at the periapsis.
We note that on an eccentric orbit, $\dot{M}/M$ may vary smoothly along the orbit. For instance, it might be reasonable to assume that $\dot{M}$ has a local maximum and minimum at periapsis and apoapsis, respectively \citep{2019ApJ...872..119H}. Detailed modelling of mass transfer in eccentric orbits is the subject of ongoing research which is beyond the scope of this paper.
Instead, we restrict ourselves to the two limiting cases (i) and (ii). We may assume that the former case is a valid approximation for circular and less eccentric systems whereas the latter holds for more eccentric orbits.

In case (i), $\dot{M}/M$ and $S$ only change on timescales that are much longer than the orbital period which is $T\lesssim\mathcal{O}(10^3)\,\text{days}$ for the systems we will be interested in (see Section~\ref{sec:population}). Hence, we can fix $\dot{M}/M$ and $\bm{S}$ when averaging Eq.~\eqref{eq:zeta-1} over one period. For that purpose, it is convenient to introduce a rotating reference frame $\mathfrak{F}$ by defining a right-handed orthonormal triad $(\bm{\hat{e}},\bm{\hat{v}},\bm{\hat{h}})$. Here, $\bm{\hat{e}}$ is the unit vector of the Laplace-Runge-Lenz vector $\bm{e}$ that has a magnitude equal to the orbit's eccentricity $e$ and points towards its periapsis. Meanwhile, the unit vector $\bm{\hat{v}}=\bm{\hat{h}}\times\bm{\hat{e}}$ is along the latus rectum of the orbit. In this frame, $\bm{\hat{d}}$ and $\bm{\hat{S}}$ read in spherical coordinates
\begin{align}
    \bm{\hat{d}}&=\cos\nu\bm{\hat{e}}+\sin\nu\bm{\hat{v}},\\
    \bm{\hat{S}}&=\cos\phi\sin\theta\bm{\hat{e}}+\sin\phi\sin\theta\bm{\hat{v}}+\cos\theta\bm{\hat{h}},
\end{align}
where $\phi\in[0,2\pi)$ and $\theta\in[0,\pi]$ are the azimuthal and polar (tilt) angles of $\bm{\hat{S}}$, respectively. The angle $\nu\in[0,2\pi)$ is the azimuthal angle of $\bm{\hat{d}}$ which is equivalent to the binary's orbital phase.

In general, the orbit-averaged change of some stellar quantity $G$ over an orbit with eccentricity $0\leq e<1$ is given by \citep{2016ApJ...825...71D}
\begin{equation}\label{eq:orbit-average}
    \langle\dot{G}(t)\rangle=\frac{(1-e^2)^{3/2}}{2\pi}\int_0^{2\pi}\frac{\dot{G}}{(1+e\cos\nu)^2}\dd\nu .
\end{equation}
For simplicity, we assume that also $e$ does not change significantly on orbital timescales so that we can set $e=\text{const.}$ in the integral of Eq.~\eqref{eq:orbit-average}. Thus, we find for the orbit-averaged change of $\bm{S}$
\begin{align}
    \langle\bm{\dot{S}}(t)\rangle=&-\frac{1}{\kappa}\frac{\dot{M}}{M}S\{\left[1-f_1(e)\right]\cos\phi\sin\theta\bm{\hat{e}}\nonumber\\
    &+\left[1-f_2(e)\right]\sin\phi\sin\theta\bm{\hat{v}}+\cos\theta\bm{\hat{h}}\}\label{eq:diff},
\end{align}
where we defined $f_1(0)=f_2(0)=1/2$, whereas for $0<e<1$ we have
\begin{align}
    f_1(e)&=\frac{e^4+2e^2\left(j-1\right) -j+1}{je^2},\\
    f_2(e)&=\frac{(e^2-1)\left(j-1\right)}{e^2},
\end{align}
whose difference is small and always negative, $0>f_2(e)-f_1(e)>-1$. In terms of the spherical coordinates of $\bm{S}$, Eq.~\eqref{eq:diff} reads
\begin{align}
    \langle\dot{\theta}\rangle&=\frac{1}{\kappa}\frac{\dot{M}}{M}\sin\theta\cos\theta\left[f_1(e)\cos^2\phi+f_2(e)\sin^2\phi\right],\label{eq:theta-dot}\\
    \langle\dot{S}\rangle&=-\frac{1}{\kappa}\frac{\dot{M}}{M}S\left\{1-\sin^2\theta\left[f_1(e)\cos^2\phi+f_2(e)\sin^2\phi\right]\right\}\label{eq:S-dot},\\
    \langle\dot{\phi}\rangle&=\frac{f_2(e)-f_1(e)}{\kappa}\frac{\dot{M}}{M}\sin\phi\cos\phi\label{eq:phi-dot}.
\end{align}
For the special case of circular, stationary orbits ($e=\langle\dot{e}\rangle=0$), the integration of Eqs.~\eqref{eq:theta-dot}--\eqref{eq:phi-dot} is particularly simple, yielding the analytical solutions
\begin{align}
    \theta&=\tan^{-1}{\left[\left(\frac{M}{M_0}\right)^{-1/2\kappa}\tan\theta_0\right]},\label{eq:theta}\\
    S&=S_0\left[\left(\frac{M}{M_0}\right)^{2/\kappa}\cos^2\theta_0+\left(\frac{M}{M_0}\right)^{1/\kappa}\sin^2\theta_0\right]^{1/2}\label{eq:magnitude},\\
    \phi&=\text{const.}\label{eq:phi},
\end{align}
where $M_0= M(t_0)$, $\theta_0=\theta(t_0)$, and $S_0= S(t_0)$ are the donor's mass, tilt angle, and spin magnitude at the onset of mass transfer, respectively. Importantly, $\theta$ and $S$ are only functions of the initial tilt angle $\theta_0$ and fractional leftover mass $0\leq M/M_0\leq1$, i.e. the fraction between the donor's current and initial masses $M$ and $M_0$, respectively. They are not explicit functions of time. Physically, this means that the details of the functional form of $\dot{M}(t)$ are irrelevant for $\theta(t)$ and $S(t)$ as long as the integrated mass loss is the same.

In Figure~\ref{fig:donor-evolution-constant-mass-rate}, we plot $\cos\theta$ and $S/S_0$ for $e=0$ as functions of $M/M_0$ and $\cos\theta_0$ revealing two essential features. Firstly, as the donor loses mass any initial tilt angle $\theta_0$ gets flipped onto the orbital plane ($\cos\theta=0$; see  {left panel}). The only two exceptions are given by $\cos\theta_0=-1$ and $+1$ for which $\cos\theta$ remains constant. However, the latter values constitute unstable equilibria since for any small deviation we do observe a flip. Moreover, in a realistic astrophysical setting these points are irrelevant because we will never start from perfect alignment of $\bm{S}$ and $\pm\bm{h}$. Most importantly, we see that the spin flip is very efficient in the sense that even moderate mass losses, e.g. $M/M_0\simeq0.75$, cause large changes of $\cos\theta$ towards zero unless $\cos\theta_0$ is very close to $-1$ or $+1$. Hence, the orbital plane is a strong dynamical attractor for the evolution of $\bm{S}$. 

Secondly, the spin magnitude $S$ of the donor gets efficiently damped down (see  {right panel}). This is even true for $\cos\theta_0=-1$ and $+1$. In fact, the closer $\cos\theta_0$ is to these values the stronger is the spin-down. Unless the mass loss is small the spin can decrease by several orders of magnitude.

Next, we investigate case (ii) in which all mass per orbit is lost at  periapsis. To this end, we introduce a mass loss rate $\dot{M}_0>0$ such that \citep{2016ApJ...825...71D}
\begin{equation}\label{eq:dirac-delta}
    \dot{M}(\nu)=\frac{\dot{M}_0}{2\pi}\delta(\nu),
\end{equation}
where $\delta(\nu)$ is the Dirac-delta distribution. In this case, Eq.~\eqref{eq:orbit-average} yields for the donor spin,
\begin{equation}
    \langle\bm{\dot{S}}\rangle=-\frac{1}{4\pi^2\kappa}\frac{(1-e^2)^{3/2}}{(1+e)^2}\frac{\dot{M}_0}{M}S\left(\sin\phi\sin\theta\bm{\hat{v}}+\cos\theta\bm{\hat{h}}\right).\label{eq:diff2}
\end{equation}
\begin{figure*}
    \centering
    \subfigure{\includegraphics[scale=0.45]{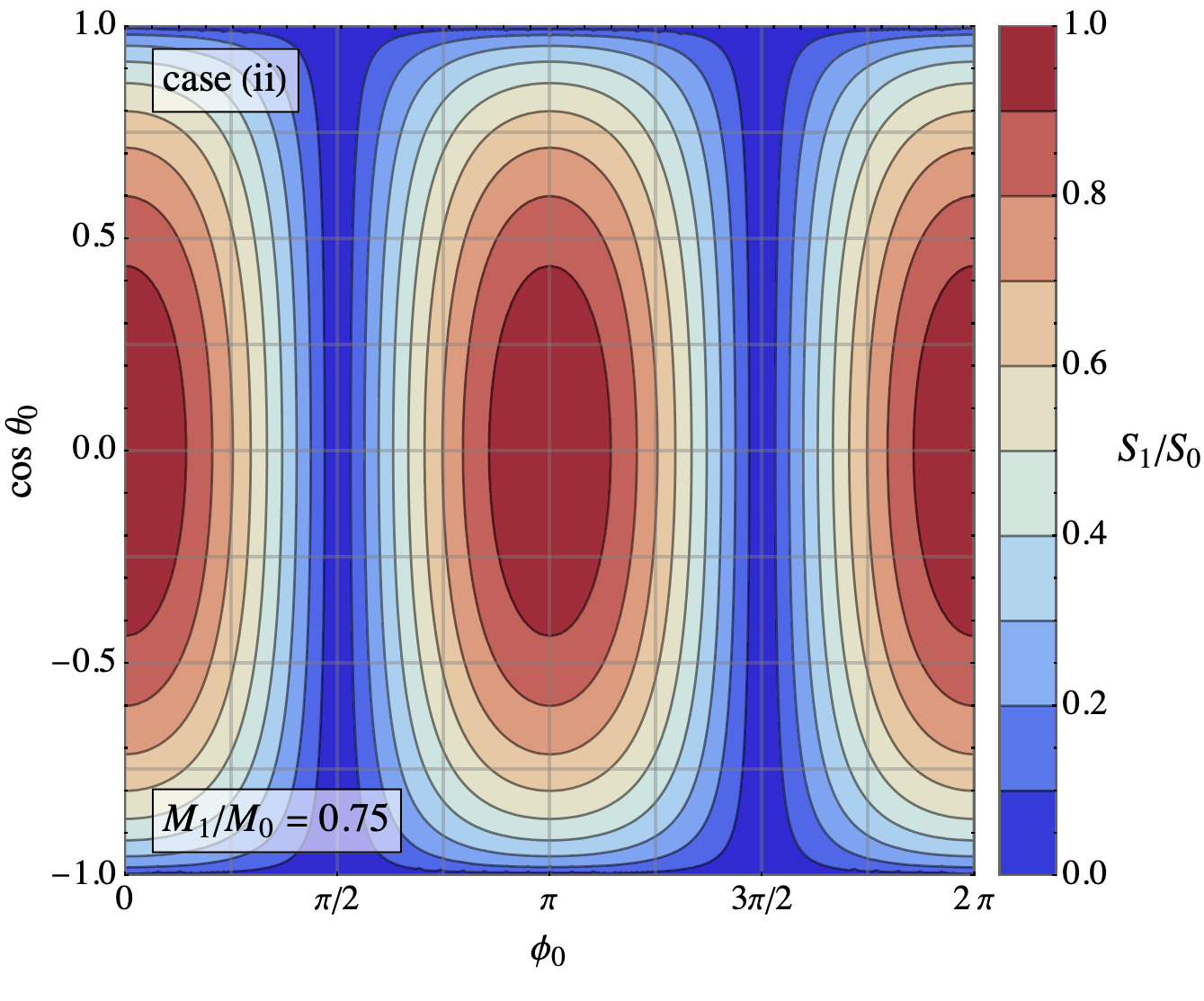}}\quad
    \subfigure{\includegraphics[scale=0.45]{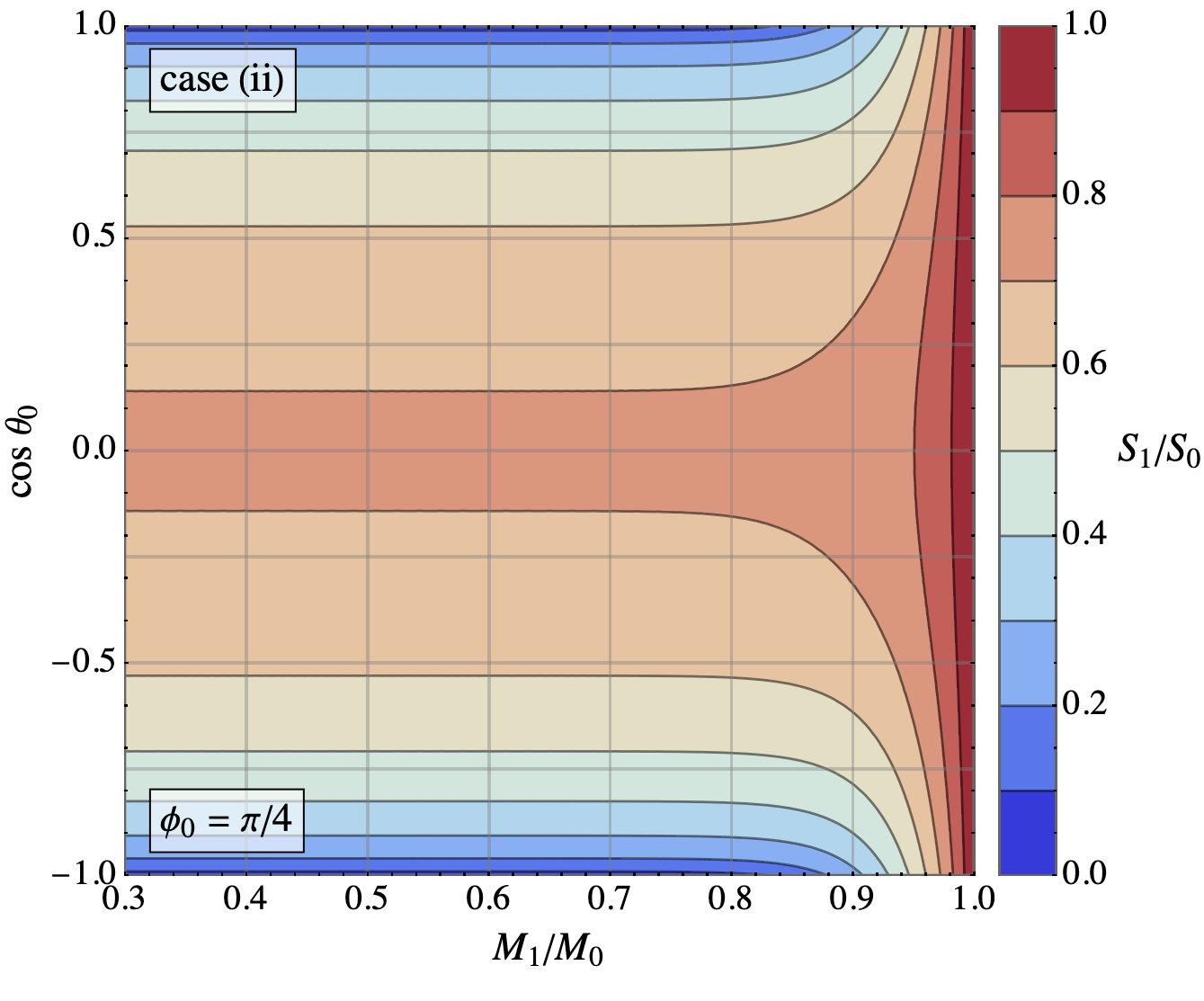}}
    \caption{Donor spin-down if all mass is lost at the periapsis. All quantities are defined as in Figures \ref{fig:donor-evolution-constant-mass-rate} and \ref{fig:donor-flip-delta-mass-rate}. For $M_1/M_0=0.75$, the  {left panel} shows the dependency on $\phi_0$. We can see that the damping of the magnitude is most (least) efficient for $\pi/2$ or $3\pi/2$ ($0$ or $\pi$) whilst the opposite was true for the spin flip (e.g., Figure~\ref{fig:donor-flip-delta-mass-rate},  {left panel}). For the fiducial angle $\phi_0=\pi/4$, the  {right panel} shows that the spin magnitude gets damped down but much less efficiently than for the circular, constant mass loss rate case (Figure~\ref{fig:donor-evolution-constant-mass-rate}).}
    \label{fig:donor-magnitude-delta-mass-rate}
\end{figure*}
In terms of the spherical coordinates we thus get
\begin{align}
    \langle\dot{\theta}\rangle&=\frac{1}{\kappa}\frac{\langle\dot{M}\rangle}{M}\cos^2\phi\sin\theta\cos\theta,\label{eq:theta-dot-delta}\\
    \langle\dot{S}\rangle&=-\frac{1}{\kappa}\frac{\langle\dot{M}\rangle}{M}S\left(\sin^2\phi\sin^2\theta+\cos^2\theta\right),\label{eq:S-dot-delta}\\
    \langle\dot{\phi}\rangle&=-\frac{1}{\kappa}\frac{\langle\dot{M}\rangle}{M}\sin\phi\cos\phi,\label{eq:phi-dot-delta}
\end{align}
where we substituted the orbit-average of Eq.~\eqref{eq:dirac-delta}
\begin{equation}\label{eq:orbit-average-mdot}
    \langle\dot{M}\rangle=\frac{\dot{M}_0}{4\pi^2}\frac{(1-e^2)^{3/2}}{(1+e)^2}.
\end{equation}
Note that since we defined $\dot{M}_0$ to be positive, the donor loses mass at a rate $-\langle\dot{M}\rangle$. Analogously to Eqs.~\eqref{eq:theta-dot}--\eqref{eq:phi}, we find analytical solutions to Eqs.~\eqref{eq:theta-dot-delta}--\eqref{eq:phi-dot-delta} which are given by
\begin{align}
    \theta&=\tan^{-1}{\left\{\left[\left(\frac{M}{M_0}\right)^{-2/\kappa}\cos^2\phi_0+\sin^2\phi_0\right]^{1/2}\tan\theta_0\right\}},\label{eq:theta-delta}\\
    S&=S_0x^\alpha y^\beta,\label{eq:magnitude-delta}\\
    \phi&=\tan^{-1}{\left[\left(\frac{M}{M_0}\right)^{1/\kappa}\tan\phi_0\right]}\label{eq:phi-delta},
\end{align}
where we defined
\begin{align}
    x=&\cos^2\phi_0\sin^2\theta_0+\left(\frac{M}{M_0}\right)^{2/\kappa}\left(\cos^2\theta_0+\sin^2\phi_0\sin^2\theta_0\right),\label{eq:x}\\
    y=&\frac{1}{4}\Bigg[1-\cos(2\theta_0)+2\cos(2\phi_0)\sin^2\theta_0\nonumber\\
    &+\left(\frac{M}{M_0}\right)^{2/\kappa}\left(3+\cos(2\theta_0)-2\cos(2\phi_0)\sin^2\theta_0\right)\Bigg],\label{eq:y}\\
    \alpha=&\frac{1}{2+2\cot^2\theta_0\csc^2\phi_0},\\
    \beta=&\frac{2\cos^2\theta_0}{3+\cos(2\theta_0)-2\cos(2\phi_0)\sin^2\theta_0}.
\end{align}

In the following, we investigate the implications of Eqs.~\eqref{eq:theta-delta}--\eqref{eq:phi-delta}. At first, note that these equations are not explicit functions of the eccentricity. Hence, their scope is only physically, but not mathematically, restricted to eccentricities that must be large enough so that Eq.~\eqref{eq:dirac-delta} provides a valid approximation for the mass loss rate. Yet Eqs.~\eqref{eq:theta-delta}--\eqref{eq:phi-delta} are more complicated to analyse than Eqs.~\eqref{eq:theta}--\eqref{eq:phi} due to their additional dependency on $\phi_0$. The dependency of the tilt angle is shown in the  {left panel} of Figure~\ref{fig:donor-flip-delta-mass-rate} where we fixed $M/M_0=0.75$. It reveals that the flip is more efficient the closer the spin starts around the periapsis or apoapsis, i.e. around $\phi_0=0$ or $\pi$, respectively, whilst it does not flip at all if $\phi_0=\pi/2$ or $3\pi/2$. The limiting behaviour at these values becomes evident from Eqs.~\eqref{eq:theta-dot-delta} and \eqref{eq:phi-dot-delta}. If $\phi_0$ is an integer or half-integer multiple of $\pi$, $\phi$ becomes stationary ($\langle\dot{\phi}\rangle=0$) since $\sin\phi\cos\phi=0$ for these cases. Therefore, $\langle\dot{\theta}\rangle$ scales with a constant factor $\cos
^2\phi_0$ which is zero or one if $\phi_0$ is an half-integer or integer multiple of $\pi$, respectively. In the  {right panel} of Figure~\ref{fig:donor-flip-delta-mass-rate}, we plot $\cos\theta$ for $\phi_0=0.49\pi$, i.e. for an initial azimuthal angle close to a half-integer minimum. We see that it requires smaller values of $M/M_0$ compared to Figure~\ref{fig:donor-evolution-constant-mass-rate} in order to achieve a significant flip. The opposite would be true if we were to plot the panel for $\phi_0$ around $0$ or $\pi$.

In Figure~\ref{fig:donor-magnitude-delta-mass-rate}, we show the spin magnitude evolution. In the  {left panel}, we show its dependency on $\phi_0$ for $M/M_0=0.75$. Whilst we explained above that the spin flip is most (least) efficient if $\phi_0$ is an integer (half-integer) multiple of $\pi$, respectively, the opposite is true for the spin-down. In the  {right panel}, we show for $\phi_0=\pi/4$ the spin-down as a function of $\cos\theta_0$. In comparison to the circular, constant mass loss rate case (Figure~\ref{fig:donor-evolution-constant-mass-rate}), the spin-down is weaker leading to a typical fraction of about $S_1/S_0\sim\mathcal{O}(0.1)$. Furthermore, there is a mass loss scale of about $M/M_0\simeq0.8$ below which the spin-down no longer depends on $M/M_0$. At this value, terms that are proportional to $(M/M_0)^{2/\kappa}$ become negligible in Eqs.~\eqref{eq:x} and \eqref{eq:y} so that $S/S_0$ solely depends on $\theta_0$ and $\phi_0$. In this regime, the spin-down is the most (least) efficient if the spin is oriented towards the poles (orbital plane).

\section{Tides}\label{sec:TR}
In a close semi-detached binary tidal bulges can emerge on the surface of a star because of the perturbing force of its companion. Due to the viscosity of the star these bulges will not instantaneously align with the relative distance vector $\bm{d}$ but they either slightly lag behind or lead ahead depending on whether its rotational angular frequency $\omega$ is smaller or greater than the orbital mean motion $n=2\pi/T$, respectively \citep{1973Ap&SS..23..459A,1977A&A....57..383Z,1981A&A....99..126H}. Further bulges arise at the star's equator due to its rotation introducing at lowest order a quadrupolar perturbation to the gravitational potential. Applying both effects to the donor star, the equations of motion for the evolution of the stellar spin and binary angular momentum are described by a
set of differential equations for $\bm{S}$, $\bm{\hat{e}}$, $\bm{\hat{v}}$, $\bm{\hat{h}}$, $e$, and $h$ as follows
\begin{align}
    \frac{\d\bm{S}}{\d t}&={\frac{\d\bm{S}}{\d t}}\Bigg|_{\dot{M}}+{\frac{\d\bm{S}}{\d t}}\Bigg|_{\text{Quad}}+{\frac{\d\bm{S}}{\d t}}\Bigg|_{\text{Diss}},\label{eq:Num-S1}\\
    \frac{\d\bm{\hat{u}}}{\d t}&={\frac{\d\bm{\hat{u}}}{\d t}}\Bigg|_{\text{Quad}}+{\frac{\d\bm{\hat{u}}}{\d t}}\Bigg|_{\text{Diss}},\label{eq:Num-u}\\
    \frac{\d e}{\d t}&={\frac{\d e}{\d t}}\Bigg|_{\dot{M}}+{\frac{\d e}{\d t}}\Bigg|_{\text{Diss}},\label{eq:Num-e}\\
    \frac{\d h}{\d t}&={\frac{\d h}{\d t}}\Bigg|_{\dot{M}}+{\frac{\d h}{\d t}}\Bigg|_{\text{Diss}}\label{eq:Num-h},
\end{align}
where $\bm{\hat{u}}$ is used as a proxy for $\bm{\hat{e}}$, $\bm{\hat{v}}$, and $\bm{\hat{h}}$, respectively.
For simplicity, we ignore tides raised on the acceptor star which are much weaker than the tides raised on the donor. 
Each term in Eqs.~\eqref{eq:Num-S1}--\eqref{eq:Num-h} either emerges from the mass transfer (indicated by "$\dot{M}$"), the quadrupolar distortion of the donor ("Quad"), or the tidal dissipation ("Diss"). The mass transfer term for $\bm{S}$ is either given by Eq.~\eqref{eq:diff} [case (i)] or \eqref{eq:diff2} [case (ii)]. In addition, conservative mass transfer causes the magnitude of the specific angular momentum to change as \citep{EggletonP.P.PeterP.2006Epib}
\begin{equation}
    {\frac{\d h}{\d t}}\Bigg|_{\dot{M}}=\dot{M}\left(\frac{1}{M_1}-\frac{1}{M_2}\right)h.
\end{equation}

Meanwhile, the mass transfer term for the eccentricity depends on how the mass loss rate changes along the orbit. If the rate is independent on the orbital phase
as in case (i), then \citep{EggletonP.P.PeterP.2006Epib} 
\begin{equation}
    {\frac{\d e}{\d t}}\Bigg|_{\dot{M}}=0,
\end{equation}
whereas for the delta-mass function case (ii), it can be written as \citep{2007ApJ...667.1170S,2016ApJ...825...71D}
\begin{equation}
    {\frac{\d e}{\d t}}\Bigg|_{\dot{M}}=-\frac{\dot{M}}{M_1}\frac{R}{a}j+2\dot{M}\left(\frac{1}{M_1}-\frac{1}{M_2}\right)j(1-e),
\end{equation}
where we treated the accretor as a point mass. In any case, the orientation of the orbital frame $\mathfrak{F}$ remains unaffected by the mass transfer \citep{EggletonP.P.PeterP.2006Epib}. Hence, Eq.~\eqref{eq:Num-u} involves no term in this regard. 

Together, the terms for the quadrupolar distortion and tidal dissipation can be conveniently expressed by means of five perturbing functions $X$, $Y$, $Z$, $V$, and $W$ which we explicate in Appendix \ref{sec:apppendix-A} \citep{2001ApJ...562.1012E,EggletonP.P.PeterP.2006Epib}
\begin{align}
    {\frac{\d\bm{S}}{\d t}}\Bigg|_{\text{Quad}}+{\frac{\d\bm{S}}{\d t}}\Bigg|_{\text{Diss}}&=\mu h(W\bm{\hat{h}}-\bm{K}\times\bm{\hat{h}}),\label{eq:diff-S}\\
    {\frac{\d\bm{\hat{u}}}{\d t}}\Bigg|_{\text{Quad}}+{\frac{\d\bm{\hat{u}}}{\d t}}\Bigg|_{\text{Diss}}&=\bm{K}\times\bm{\hat{u}},\label{eq:Angular-L}\\
    {\frac{\d e}{\d t}}\Bigg|_{\text{Diss}}&=-Ve,\label{eq:diff-e}\\
    {\frac{\d h}{\d t}}\Bigg|_{\text{Diss}}&=-Wh,\label{eq:diff-h}
\end{align}
where $\bm{K}= X\bm{\hat{e}}+Y\bm{\hat{v}}+Z\bm{\hat{h}}$ is the angular velocity of $\mathfrak{F}$. Of the perturbing functions, $V$ and $W$ are due to tidal dissipation which cause the orbit to circularise and the stellar rotation to synchronise. The functions $X$, $Y$, and $Z$ incorporate the quadrupolar distortion which gives rise to apsidal motion and spin precession around $\bm{\hat{h}}$. However, also $X$ and $Y$ do include small terms due to tidal dissipation which enforce the spin to align with the orbital angular momentum. Hence, the effect of tidal dissipation counteracts the flip of the donor spin due to mass transfer, i.e. its misalignment with $\bm{\hat{h}}$. Therefore, any spin flip is suppressed unless the mass transfer terms in Eq.~\eqref{eq:Num-S1} are able to dominate the others. 
In the following, we address in which circumstances this might happen.

In the equilibrium tide model the tidal friction timescale $t_\text{F}$ is defined as \citep{2007ApJ...669.1298F}
\begin{equation}
    t_\text{F,e}=\frac{t_\text{V}}{9}\left(\frac{a}{R}\right)^8\frac{M_1^2}{M_2M_{12}}\frac{1}{(1+2k_{\rm A})^2},
\end{equation}
where $k_{\rm A}=0.014$ is the apsidal motion constant which quantifies the quadrupolar deformability of a star and $t_\text{V}$ is the viscous timescale given by
\begin{equation}\label{eq:tvis}
    t_\text{V}=3\frac{(1+2k_A)^2}{k_A}\frac{R^3}{GM_1\tau}.
\end{equation}
 Physically, $\tau$ describes the time by which the tidal bulges lag behind or lead ahead w.r.t. the line connecting both binary members. In the
 theory of equilibrium tides, $\tau$ is a constant which is an intrinsic property of the tidally forced star in question \citep{1981A&A....99..126H}. 

We also consider an approximate prescription for dynamical tides which could become important for stars with outer radiative envelopes \citep{1977A&A....57..383Z}. Following \citet{2002MNRAS.329..897H}, in this case
we still use the equations from \citet{2001ApJ...562.1012E} 
but with 
the tidal dissipation timescale now given by
\begin{equation}
    t_\text{F,d}=\left(\frac{a}{R}\right)^9\sqrt{\frac{a^3}{GM_1}} q \left(1+{1\over q} \right)^{-11/6} \frac{1}{E_2},
\end{equation}
where $E_2$ is a coefficient that is related to the structure of the
star and refers to the coupling between the tidal potential and
gravity mode oscillations.
Unfortunately, the value of $E_2$ is difficult to calculate since it is very
sensitive to the structure of the star and therefore to the exact treatment of stellar evolution \citep[e.g.,][]{1975A&A....41..329Z,2010ApJ...725..940Y,2013A&A...550A.100S,2018A&A...616A..28Q}. 

We can approximately quantify the effect of tides on the stellar spin
by introducing a timescale
for variations in the spin-orbit tilt angle
due to tidal dissipation, $t_{\rm S,e(d)} 
\sim {(S/\mu h)}t_{\rm F, e(d)} j^{13}$.
Thus, for a star which just fills its Roche lobe, i.e., $R=F(q)a(1-e)$, and for the equilibrium tide model we have
\begin{eqnarray}
t_{ S,{\rm e}}=&&
{1.6\times 10^{-2} \,\text{yr}}  {\kappa\over0.08}{0.014\over k_{\rm A}}
 \left( {1\,\rm s}\over \tau \right)
\left( {1\,\rm day} \over P \right) 
 \nonumber \\  &&\times \left({\msun^{3/2}\over{M_2\sqrt{M_{12}}}} \right) \left(a\over \rsun \right)^{9/2} 
{qj^{12}\over \left[ F(q)(1-e) \right]^3 },
\end{eqnarray}
where $P=2\pi/\omega$.
For the dynamical tide model we find
\begin{eqnarray}
t_{ S,{\rm d}}=&&
{4.4\times 10^2 \,\text{yr}} {\kappa\over0.08}\left({10^{-9}\over E_2}\right)
\left( {1\,\rm day} \over P \right) 
\nonumber \\  &&\times \left( {\sqrt{M_1M_{12}}\over M_2^2}
{ \msun}\right) 
\left(a\over \rsun\right)^{3} \nonumber \\ &&\times {j^{12}\over \left[ F(q)(1-e) \right]^7 }\left(M_{2}\over M_{12} \right)^{11/6}.
\end{eqnarray}
 We then compare the above timescales 
to the  timescale for spin change due to mass loss,
$t_{ \dot{M}}\sim \kappa M_1/\dot{M}$. 
From the condition $t_{\dot{M}}=t_{S,{\rm e(d)}}$, we have that the mass loss effect dominates over tidal effects
if at the onset of mass transfer the binary semi-major axis is larger than
\begin{eqnarray}
a_{\rm e}=&&11.9\rsun\left({k_{\rm A}\over 0.014}
{\tau \over  1\,\rm s}
{10^{-4}\msun{\,\text{yr}^{-1}} \over {\dot{M}}}
{P \over  1\,\rm day }
\right)^{2/9} \nonumber \\
&&\times
\left({M_2^2 \sqrt{M_{12}} \over \msun^{5/2}}\right)^{2/9}
{\left[ F(q)(1-e) \right]^{2/3} \over j^{8/3}},
\end{eqnarray}
for equilibrium tides, and larger than
\begin{eqnarray}
a_{\rm d}=&&1.2\rsun\left(\frac{E_2}{10^{-9}}
{\frac{10^{-4}\msun{\,\text{yr}^{-1}}}{\dot{M}}}
{\frac{P}{1\,\rm day}}
{\frac{M_2^2 M_1}{\sqrt{M_1M_{12}}\msun^2}}\right)^{1/3}
\nonumber\\
&&\times\frac{\left[F(q)(1-e)\right]^{7/3}}{j^4}\left(\frac{M_{12}}{M_{2}}\right)^{11/18},
\end{eqnarray}
for dynamical tides.

\begin{figure}
\centering
\includegraphics[width=0.45\textwidth]{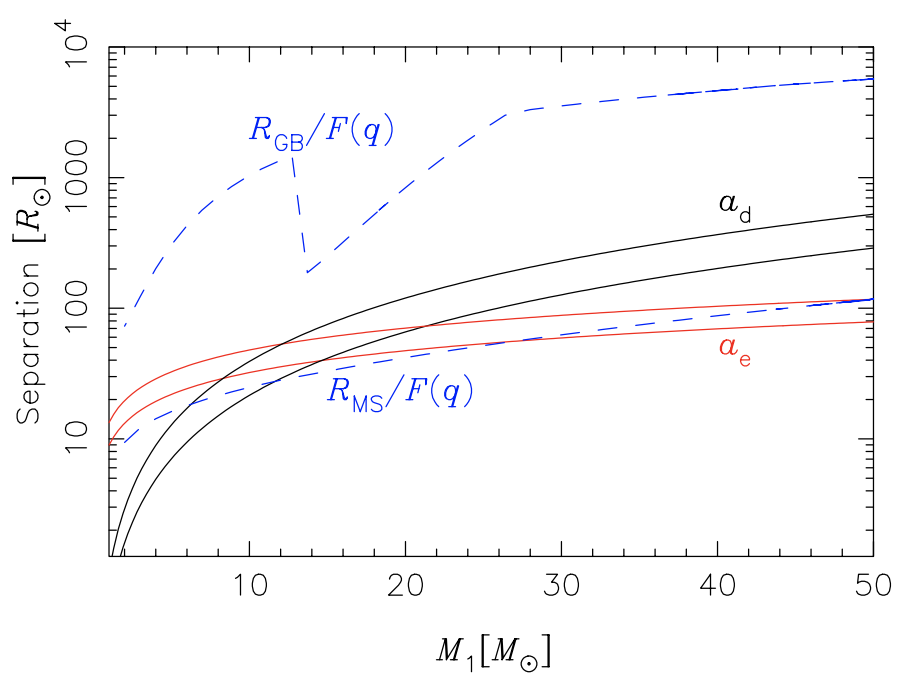}
\caption{Critical binary semi-major axis $a_{\rm e}$ and $a_{\rm d}$ for a spin rotational period $P=5\,\rm days$ (lower lines) and 
$P=30\,\rm days$ (upper lines).  Tides are expected to suppress the flip of the stellar spin due to mass loss if mass transfer starts at
$a<a_{\rm e}$ and $a<a_{\rm d}$ for equilibrium and dynamical tides, respectively. The dashed blue lines give the maximum value of the binary semi-major axis that will still allow a mass transfer event to occur on the main-sequence, $R_{\rm MS}/F(q)$, and before He burning starts, $R_{\rm GB}/F(q)$. 
Here, we consider circular binaries with $e=0$, $q=1$, and solar metallicity. Other parameters and details are given in Section \ref{sec:TR}.
}
\label{fig:aed}
\end{figure}

In Fig.~\ref{fig:aed}, we plot $a_{\rm e}$
and $a_{\rm d}$ as a function
of the donor mass. In this calculation, we consider circular equal-mass binaries.
 For the lag time constant we use $\tau=10^{-1}\,\text{s}$ which is a value typically adopted for solar type stars
 \citep[e.g.,][]{2003ApJ...589..605W,2017MNRAS.467.3066A}, and set $E_2= 10^{-9} \left({M_1/\msun}\right)^{2.8}$ which was
obtained by \citet{1977A&A....57..383Z} for zero-age main-sequence stars.
 Although  $a_{\rm e}$
and $a_{\rm d}$ depend weekly on 
 $\tau$ and $E_2$ respectively, it is important to note that plausible values for these parameters can span orders of magnitude \citep[e.g.,][]{1966Icar....5..375G,1977A&A....57..383Z}.
Moreover, we set
${\dot{M}}=10^{-4}\msun{\,\text{yr}^{-1}}$ which is a realistic value for
Roche lobe overflow.
According to \citet{1967ZA.....66...58K} a $9\msun$ donor loses more than
$5\msun$ to a $5\msun$ accretor in only
$6\times 10^4\,\text{yr}$ during the hydrogen burning
and almost $7\msun$ in $4\times 10^4\,\text{yr}$
when the mass transfer starts after exhaustion of hydrogen in the core. \citet{1967AcA....17..193P} and \citet{2010A&A...510A..13V} find similar mass transfer rates for  main-sequence donors. The mass loss rate can be as large as $10^{-1}\,\msun\,\text{yr}^{-1}$ in the case of massive binaries \citep[e.g., Figure 1 of][]{2016Natur.534..512B}.

Figure~\ref{fig:aed}
shows that $a_{\rm e}$
varies between several tens to $\sim 100\rsun$,  
increasing weakly with the mass of the donor, whilst
$a_{\rm d}$ varies between 
 $a_{\rm d}\simeq 2\rsun$ for  a $2\msun$ donor
 up to  
$a_{\rm d}\simeq 400\rsun$ for the most massive stars.
We can now ask whether during a given evolutionary stage 
a mass transfer episode can occur at such, or smaller, orbital separations; this requires that $R/F(q)(1-e)>a_{\rm e(d)}$.
In Figure~\ref{fig:aed} we compute $R_{\rm MS}/F(q)$ and
$R_{\rm RG}/F(q)$, where $R_{\rm MS}$ is the maximum stellar radius  during the main-sequence and 
$R_{\rm GB}$ is the radius at the start of He burning, as a function of the mass of the star at that evolutionary stage. The stellar radii
were obtained with the
fast binary stellar evolution code {\fontfamily{qcr}\selectfont BSE} \citep{2002MNRAS.329..897H,2020A&A...639A..41B}.
Hence, $R_{\rm MS}/F(q)$ and
$R_{\rm RG}/F(q)$ represent the maximum value of the binary semi-major axis that will still allow a mass transfer event to happen  during the main-sequence or before He burning starts, respectively.
Comparing these to the red and black lines in the figure, we see that the effect of mass loss can indeed be important for both main-sequence and giant stars. If mass transfer starts near the tip of the giant branch, then the mass loss effect will dominate regardless of the exact treatment of tides. Even for main-sequence stars, however, the mass loss effect can become comparable to tides for a large range of masses and dominates in some cases. For example, 
radiative damping on the dynamical tide is expected to be the most efficient dissipative mechanism  in main-sequence stars with  $M_1\gtrsim 1.5\msun$. For these
stars and for masses up to $M_1\approx 5\msun$, $R_{\rm MS}/F(q)\gg a_{\rm d}$  so that mass loss effects will dominate.

In conclusion, the results shown in this section demonstrate  that the assumption  that tides 
 will erase any spin-orbit misalignment might not  always  be valid. In the following section, we will consider how the spin dynamics we described  above  can affect the spin-orbit alignment of BBHs formed from the
evolution of field binaries.

\section{Application to black hole binary formation}\label{sec:population}
In this section, we simulate a population of isolated massive stellar binaries and use Eqs.~\eqref{eq:Num-S1}--\eqref{eq:Num-h} to investigate 
the implications of the spin dynamics described above. The population is set up using the parameter distributions that \citet[]{2012Sci...337..444S} inferred from the observation of $71$ Galactic binaries. That is, we draw the primary masses, mass ratios, orbital periods, and eccentricities from the distribution functions given in Table \ref{tab:Sana}. In the remainder of this section, we will refer to the star (and its compact remnant) that was initially the more massive one as the primary and to its companion as the secondary. 

\begin{table*}
\caption{Initial parameter distribution of the binary population. The primary mass is drawn from the \citet[]{2002Sci...295...82K} initial mass function whilst the distributions for the mass ratio, eccentricity, and orbital period are adopted from \citet[]{2012Sci...337..444S}. Note that the observational sample used by \citet[]{2012Sci...337..444S} only allowed for a statistical analysis of binaries with primary masses up to $60\msun$. Here, we explicitly assume that the distributions are valid up to primary masses of $100\msun$. The exponents $\varkappa$, $\eta$, and $\lambda$ are assumed to follow normal distributions, i.e. for each binary instance we draw new values from normal distributions with means and standard deviations as given in the Table.} \label{tab:Sana} 
\begin{ruledtabular}
\begin{tabular}{lccc}
  Parameter & Distribution & Exponent & Domain\\ \hline
Primary mass $M_1$ & $P_{M_1}\propto M_1^\varsigma$ & $\varsigma=-2.3$ & $22-100\msun$\\ 
Mass ratio $M_2/M_1$ & $P_{M_2/M_1}\propto {(M_2/M_1)}^\varkappa$ & $\varkappa=-0.2\pm0.6$ & $0.1-1.0$\\ 
Eccentricity $e$ & $P_e\propto e^\eta$ & $\eta=-0.4\pm0.2$ & $0.0-0.9$\\
Orbital period $T$ & $P_{\log  T/\text{days}}\propto(\log  T/\text{days})^\lambda$ & $\lambda=-0.4\pm0.2$ & $10^{0.15}-10^{3.5}\,$days \\
 \end{tabular}
 \end{ruledtabular}
\end{table*}

We evolve this binary population in time by means of the latest version of the binary stellar evolution code {\fontfamily{qcr}\selectfont BSE} \citep{2002MNRAS.329..897H,2020A&A...639A..41B}. {\fontfamily{qcr}\selectfont BSE} simulates the stellar evolution including binary features such as mass transfer, mass accretion, common-envelope evolution, supernova kicks, and angular momentum losses. For the subset of systems that form merging BBHs through the isolated binary channel, interactions among the stellar binary members played a vital role \citep[e.g.,][]{2016Natur.534..512B}. Briefly, starting with two massive stars in the Galactic field the primary star transfers mass to the secondary during a dynamically stable Roche lobe overflow phase. Soon after this process the primary star forms a (first-born) black hole whereas the secondary expands as a supergiant. A second mass transfer phase from the secondary star to the black hole takes place once the former fills its Roche lobe. This time, the process is dynamically unstable leading to a common-envelope phase in which the expanding star engulfs its black hole companion. Whilst moving through the common envelope, drag forces cause the black hole's orbit quickly to shrink and the common envelope might be ejected. Eventually, a BBH forms once the secondary star develops a black hole, too. On this evolutionary pathway it is the common-envelope phase which is of crucial importance for the BBH to finally merge. It can rapidly reduce the orbital separation to values small enough for energy loss due to gravitational wave emission to provoke a coalescence within the age of the Universe.

In total, Eqs.~\eqref{eq:Num-S1}--\eqref{eq:Num-h} constitute a set of fourteen coupled differential equations (vectorial quantities counting thrice) that we numerically integrate once Roche lobe overflow starts. At that time, we use the masses, orbital parameters, and spin rates computed by {\fontfamily{qcr}\selectfont BSE} as the initial values for our integration scheme. Furthermore, we draw the initial value $\phi_0$ for the azimuthal angle of the donor from a uniform distribution between zero and $2\pi$. For the initial value of $\cos\theta_0$ of the tilt angle, we assume a uniform distribution in the interval $[0.9,1.0)$. Thus, our approach is conservative in the sense that we start with donor spins which are fairly aligned with the orbital angular momentum. During the integration, we follow \citet{2002MNRAS.329..897H} by using $R=a(1-e)F(q)$ as the effective donor radius. For comparison, we also ran a simulation using $R=aF(q)$ without noticing a substantial difference of the results. The mass loss rate we adopted is also obtained from the {\fontfamily{qcr}\selectfont BSE} calculation. This is, however, based on the assumption that the binary moves on a circular orbit which is not always the case for our binaries. Nonetheless, for want of a more detailed treatment we assume that this mass loss rate is still applicable to our eccentric systems. In particular, the {\fontfamily{qcr}\selectfont BSE} mass loss rate is used for the orbit-average given by Eq.~\eqref{eq:orbit-average-mdot} in case (ii).

Because the physics of stellar
tides is much debated and the efficiency of tides itself is  uncertain \citep[e.g.,][]{1997A&A...318..187C,2009A&A...500..133L}, in the simulations presented here we opt for a simplified approach in which we employ the equilibrium tide equations for all stars. Then we use the constant time lag as a free parameter  in order to tune the  efficiency of tides. We set $\tau =10^0\,\text{s}$ (efficient tides), $\tau =10^{-1}\,\text{s}$ (moderately efficient tides), and $\tau =10^{-2}\,\text{s}$ (inefficient tides). For $M_1=50\msun$, $R=10\rsun{}$, and $\tau=10^0\,\text{s}$, Eq.~\eqref{eq:tvis} gives a viscous time $t_\text{V}\simeq360\,\text{yr}$.

Before presenting the results of our analysis, we comment on some assumptions in our treatment that we briefly introduced in Section~\ref{sec:assumptions} and that are also commonly adopted in the literature. Firstly, because the extent to which the rotation of the stellar core is coupled to that of the stellar envelope is very uncertain, we simply assume maximal coupling, i.e., that the entire star behaves as a rigid rotator with a uniform angular velocity \citep{2020arXiv201000078S,2010ApJ...724..546S,2008ApJS..174..223B,2000MNRAS.315..543H,2020A&A...636A.104B}, but comment here on the other extreme case of minimal coupling in which core and envelope are fully decoupled.
As long as a star remains homogeneous, various processes (e.g., shear instability)  will tend to rapidly restore uniform rotation.
Thus, when the stars are on the main-sequence the assumption of solid rotation might represent a good approximation. But, once the star leaves the main-sequence it will then  develop a compact He rich core whose rotation could fully decouple from that of the envelope before any significant amount of mass has been lost by the donor.
The validity of our treatment for post main-sequence stars therefore requires that mass and angular momentum transport within the star are efficient enough that the stellar core remains strongly coupled to the outer envelope. The angular momentum evolution of stellar interiors, along with the resulting rotation rates of stellar remnants, remains poorly understood. However, several  studies have shown that angular momentum transport within massive stars might be efficient enough to carry a significant amount of spin from the core to the envelope \citep[e.g.,][]{2018A&A...616A..28Q,2019ApJ...881L...1F}.
In this case, a spin tilt predicted by our model will reflect onto the spin of the core as well, although the latter might still rotate at a somewhat different angular frequency and at a different angle than the envelope. If the core and envelope are fully decoupled, we would expect that  the core will keep rotating in the same direction as the entire star at the onset of mass transfer. The relative orientation between the spins of the binary stars and their orbital angular momentum then will largely depend on whether tides were efficient enough to realign any prior spin-orbit tilt, and on the primordial spin-orbit alignment. 

Secondly, following \citet{2007ApJ...667.1170S,2010ApJ...724..546S} and \citet{2016ApJ...825...71D,2016ApJ...825...70D} we assume that any orbital angular momentum carried by the loss particles is immediately returned (only) to the orbit once they have passed the first Lagrangian point. Generally, mass transfer becomes non-conservative if not all mass lost from the donor can be accreted by its companion \citep[][]{2012IAUS..282..417T}. In this case, the systemic mass and angular momentum losses would change the orbital elements differently compared to a conservative mass transfer. \citet[]{2009ApJ...702.1387S} showed for the case where all mass per orbit is lost at periapsis that the orbit would expand (contract) faster (slower). They found the same tendency for the growth (damping) of the eccentricity. Meanwhile, we showed in Section~\ref{sec:spin-evolution} that the donor spin dynamics foremostly depends on the donor's fractional mass loss rate. Thus, we expect our results to change under the consideration of non-conservative mass transfer only if the orbital elements are able to alter the latter significantly compared to the conservative case.

\begin{figure}
\centering
\includegraphics[width=0.47\textwidth]{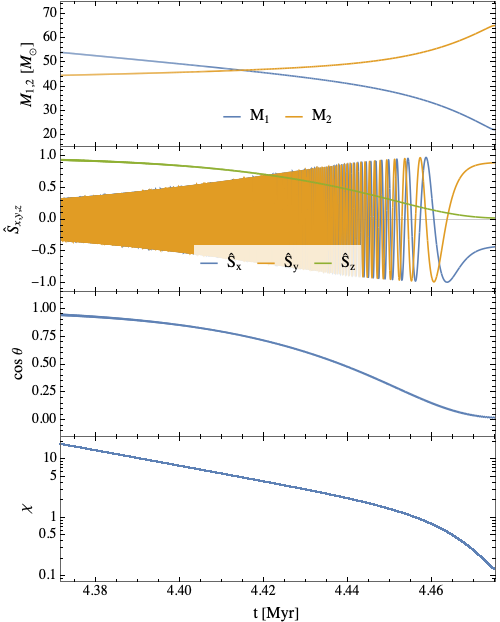}
\caption{Spin-evolution in one exemplary mass transferring stellar binary. At $t=0$ (ZAMS), the stellar masses, eccentricity, and orbital period were set to $M_1=55\msun$, $M_2=45\msun$, $e=0.1$, and $T=10^2\,\text{days}$, respectively. This binary undergoes a mass transfer phase at about $4.37$ to $4.48\,\text{Myr}$ which is considered here. The tidal lag time was set to $\tau=10^{-1}\,\text{s}$. From top to bottom, the  {first panel} shows the evolution of the stellar masses, the  {second panel} the donor spin components in some inertial frame, the  {third panel} the cosine of the tilt angle, and the  {fourth panel} its dimensionless spin magnitude $\chi=cS/GM_1^2$. The inertial frame is defined such that it coincides with $\mathfrak{F}$ at the onset of mass transfer (i.e. the $z$-axis is pointing along $\bm{\hat{h}}$).}
\label{fig:single-evolution}
\end{figure}

In the following, we present the results of our analysis. As a typical example, we show in Figure~\ref{fig:single-evolution} the spin flip during the mass transfer phase of a binary at low metallicity ($Z=0.03\,\zsun$) modelled with case (i) and a tidal lag time of $\tau=10^{-1}\,\text{s}$. This system started on the zero-age-main-sequence (ZAMS, $t=0$) with stellar masses, eccentricity, and orbital period set to $M_1=55\msun$, $M_2=45\msun$, $e=0.1$, and $T=10^2\,\text{days}$, respectively. The mass transfer phase in question lasts from $4.37$ to $4.48\,\text{Myr}$. In this period of time, the donor transfers about $60$ per cent of its mass to the accretor ({first panel}). The normalised components of the donor spin in some inertial frame are shown in the  {second panel}. This inertial frame is chosen such that the $z$-axis initially points along $\bm{\hat{h}}$. However, even at later time their directions will not deviate significantly from one another since $\bm{K}$ stays almost parallel to $\bm{\hat{h}}$ during the process. Thus, the reduction of $\hat{S}_z$ from about one to zero indicates the flip onto the orbital plane which can be also directly seen by the evolution of the cosine of the tilt angle, $\cos\theta=\bm{\hat{S}}\cdot\bm{\hat{h}}$ ({third panel}). The oscillations of the other two components describe the spin precession around $\bm{\hat{h}}$. Meanwhile, in terms of the dimensionless parameter $\chi=cS/GM_1^2$ the  {fourth panel} shows that the donor spin magnitude decreases by approximately two orders of magnitude ending up at a value $\chi\sim\mathcal{O}(0.1)$. Evidently, the value of $\chi$ at the onset and hence also at the end of the mass transfer phase depend on its initial value $\chi(t=0)$ and any torques which affect the spin until onset. Here, the initial spin is determined by a fit to the rotational velocities of main-sequence star data \citep[][]{1992adps.book.....L} following \citet[]{2000MNRAS.315..543H}. Starting with that value, {\tt BSE} computes the subsequent spin evolution taking angular momentum losses due the isotropic stellar winds and tidal interactions with the companion into account (cf. Section~\ref{sec:TR}).

After a second mass transfer phase from $M_2$ to $M_1$ starting at $5.76\,\text{Myr}$ which leads to a common-envelope evolution, this system evolves to a BBH at $6.12\,\text{Myr}$ that merges after $\sim6\,\text{Gyr}$ due to the emission of gravitational waves.

\begin{figure*}
\centering
\includegraphics[width=0.9\textwidth]{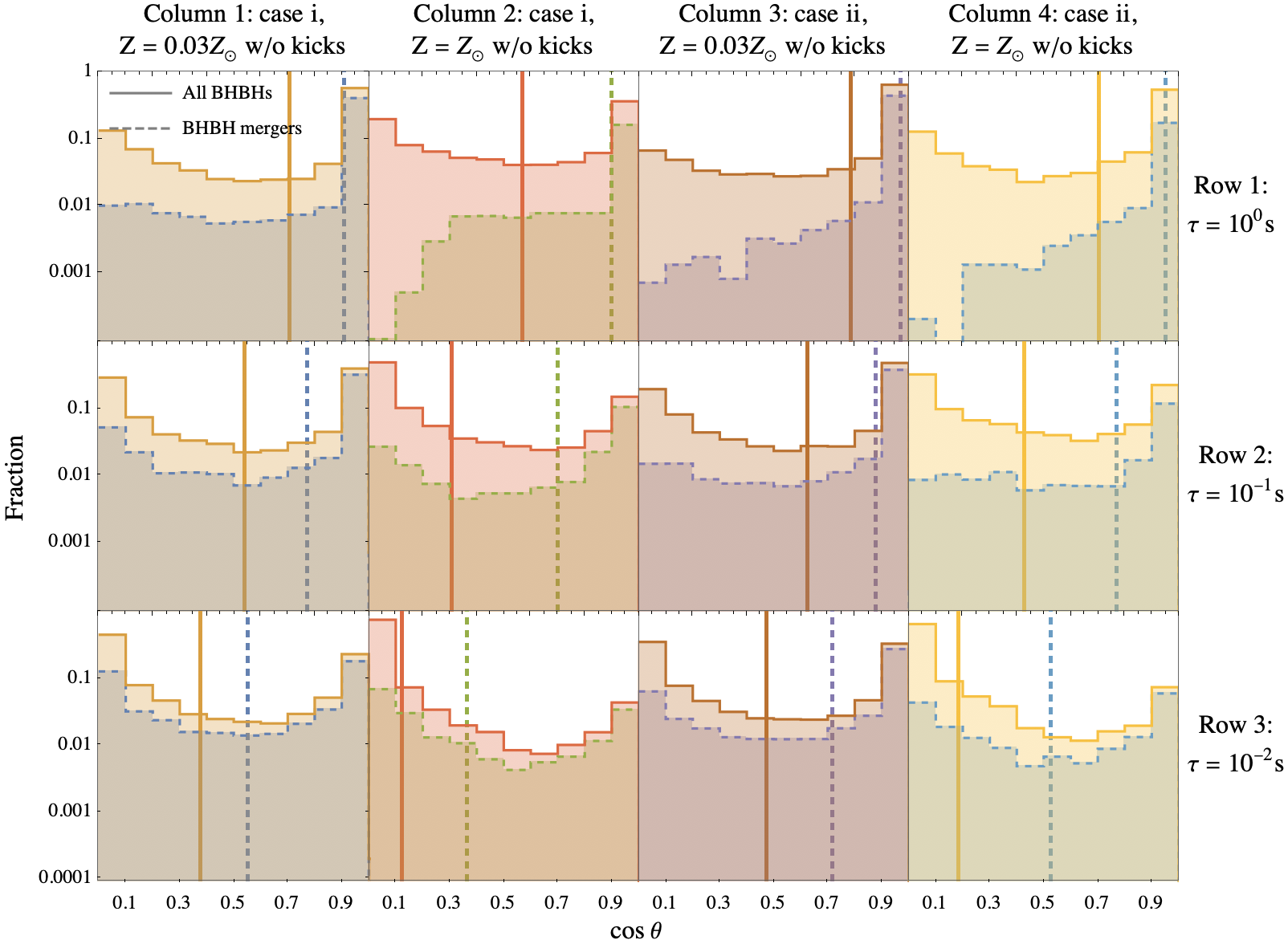}
\caption{Tilt angle distributions of the primaries once they stopped donating mass. Initially, the tilt angles were uniformly distributed in the rightmost bins ($\cos\theta=0.9-1.0$). Each solid histograms includes $10^4$ systems in which the primary fills its Roche lobe first and that eventually end up as BBH systems. The dashed histograms constitute the subset of systems that merge within $t_\text{coal}<10\,\text{Gyr}$. Vertical lines indicate the means of each distribution. From the top to the lowest panel row, the strength of the tides were decreased by lowering the lag time $\tau$. The different panel columns incorporate low and high metallicities and mass transfer modelled with case (i) and (ii) (e.g., Section~\ref{sec:spin-evolution}). }
\label{fig:histo-without-kicks}
\end{figure*}

In Figure~\ref{fig:histo-without-kicks}, we show  the primary star tilt angles for the whole binary population considering only those systems that according to {\fontfamily{qcr}\selectfont BSE} eventually develop BBHs. For each system, the value of the tilt angle is taken once the primaries stop transferring mass to their secondaries. If there was no mass transfer from the former to the latter we drew a random value from the initial value distribution of $\cos\theta_0$ between $0.9$ and $1.0$ (see above). Each histogram drawn with a solid line comprises $10^4$ systems either at low ($Z=0.03\,\zsun$) or high metallicity ($Z=\zsun$), whose lag time $\tau$ is either set to $10^0$, $10^{-1}$, or $10^{-2}\,\text{s}$, and whose mass transfer is either modelled with case (i) or case (ii). We emphasise that the bin width of $0.1$ is chosen such that the rightmost bin ($\cos\theta=0.9-1.0$) covers the range of initial angles. Thus, spins ending up in any other bin ($\cos\theta<0.9$) were dominated by the mass transfer terms. As a results, we find that any distribution is strongly bimodal with most of the primary spins either being flipped ($\cos\theta\lesssim0.1$) or remaining aligned ($\cos\theta\gtrsim0.9$). From longer to shorter lag times tides become weaker and allow the balance between the two peaks to pivot from the majority of systems being aligned to flipped. Thus, the means of the distributions indicated by the vertical solid lines shift from about $\cos\theta\sim0.7$ to $0.8$ ($\tau=10
^0\,\text{s}$) to $\sim0.5$ to $0.6$ ($\tau=10
^{-1}\,\text{s}$) and $\sim0.4$ to $0.5$ ($\tau=10
^{-2}\,\text{s}$) for low metallicity. For high metallicity, the spin flips are more effective ranging from $\cos\theta\sim0.5$ to $0.7$ ($\tau=10
^0\,\text{s}$) to $\sim0.3$ to $0.4$ ($\tau=10
^{-1}\,\text{s}$) and $\sim0.1$ to $0.2$ ($\tau=10
^{-2}\,\text{s}$). In reality, the relative number of flipped spins therefore depends on the precise value of the lag time. In turn, we do not see any major difference between our two models (i) and (ii) of mass transfer. Even though we pointed out that these models are arguably approximate with the caveats given above, the latter fact indicates that details about the orbital phase dependency of mass transfer might not play an important role for the effect that we are investigating.

In general, flipping spins are less numerous in the subsets of binaries whose black hole remnants would coalesce within $10\,\text{Gyr}$ (near the peak of cosmic star-formation rate, e.g., \citep[][]{2014ARA&A..52..415M}) taken as a rough criterion for observability by gravitational wave detectors. Yet the following results are not very sensitive to the precise numerical value. The time to coalescence can be estimated as \citep{PhysRev.136.B1224,2020MNRAS.495.2321Z}
\begin{equation}
    t_\text{coal}\simeq10\,\text{Myr}\,\left(\frac{T}{1\,\text{hr}}\right)^{8/3}\left(\frac{\msun}{M_{12}}\right)^{2/3}\left(\frac{\msun}{\mu}\right)j^7,
\end{equation}
which we evaluate once both black holes are formed. In Figure~\ref{fig:histo-without-kicks}, these subsets are indicated by dashed lines. We also display the means of the distributions by means of the vertical dotted lines. For $\tau=10^0\,\text{s}$, the number of flipped spins is insignificantly low and the means are close to or within the initial value range $[0.9,1.0)$. Hence, it can be concluded that mass transfer would be irrelevant for the spin dynamics in that case. Only if shorter lag times are considered the number of flipped spins can become comparable to that of aligned spins. In terms of the means we achieve about $0.6$ to $0.9$ and $0.3$ to $0.7$ for $\tau=10^{-1}\,\text{s}$ and $\tau=10^{-2}\,\text{s}$, respectively. The fact that flips are less prevalent in the merger subsets is due to the shorter binary separation which increases the strength of tides (see Section~\ref{sec:TR}).

Once the primary stops transferring mass to the secondary, there are four successive evolutionary stages in which its spin direction relative to $\bm{\hat{h}}$ (or that of its black hole remnant) could, in principle, change again [e.g., Figure 1 of \citet[]{2016Natur.534..512B}]. Firstly, during the rest of its lifetime the primary star is still subject to tidal forces by its companion. However, the tidal friction timescales of systems whose primaries flipped during mass transfer is typically much larger than the remaining lifetime which is about $\mathcal{O}(0.1)\,\text{Myr}$. Hence, we do not expect a significant change of the spin distributions in Figure~\ref{fig:histo-without-kicks}. Secondly, when the primary forms a black hole in a supernova the latter can receive a kick due to asymmetric mass loss that tilts the orbital angular momentum inducing a misalignment with respect to the spin directions \citep{2000ApJ...541..319K,2013PhRvD..87j4028G}. We investigate this possibility below. Thirdly, if the secondary star fills its Roche lobe it transfers mass towards the first-born black hole that has been formed out of the primary. The timescale at which the black hole spin would align with the angular momentum of an accretion disk has been derived by \citet[]{1998ApJ...506L..97N} and is given by
\begin{eqnarray}
    t_\text{align}\simeq&&0.56\,\text{Myr}\,\chi^{11/16}\left(\frac{\alpha}{0.03}\right)^{13/8}\left(\frac{L}{0.1L_\text{E}}\right)^{-7/8}\nonumber\\
    &&\times\left(\frac{M_1}{10^8\msun}\right)^{-1/16}\left(\frac{\epsilon}{0.3}\right)^{7/8},
\end{eqnarray}
where $\alpha$ the dimensionless viscosity parameter of the accretion disk \citep{1973A&A....24..337S}, $L$ the energy accretion rate onto the black hole, $L_\text{E}=1.4\times10^{38}M_1\msun^{-1}\,\text{ergs}\,\text{s}^{-1}$ the Eddington luminosity, and $\epsilon=L/\dot{M}_1c^2$ the efficiency of the accretion process. As an order-of-magnitude estimate we would get $t_\text{align}\simeq1.4\,\text{Myr}$ for $\chi=1$, $\alpha=0.03$, $L=0.1L_\text{E}$, $M_1=50\msun$, and $\epsilon=0.3$. Again, this is typically much longer than the duration of the second mass transfer phase. Fourthly, if this mass transfer is succeeded by a common-envelope phase in which the expanding envelope of the secondary engulfs the primary black hole the latter is subject to dynamical friction forces promoting a quick inspiral. Recent hydrodynamic simulations show that whilst inspiralling the mass and dimensionless spin parameter of the black hole do increase but not larger than $1$ to $2$ per cent and 0.05, respectively \citep{2020ApJ...897..130D}.

Based on this discussion, we have reason to believe that, unless natal kicks are considered, the tilt angle of the primary spin and the orbital angular momentum does not significantly change between the end of the first mass transfer phase and the formation of the black hole binary. Figure~\ref{fig:histo-without-kicks} would hence reflect the spin distributions of the first-born black hole in the BBH unless the effect of natal kicks are considered. 

\begin{figure*}
\centering
\includegraphics[width=0.9\textwidth]{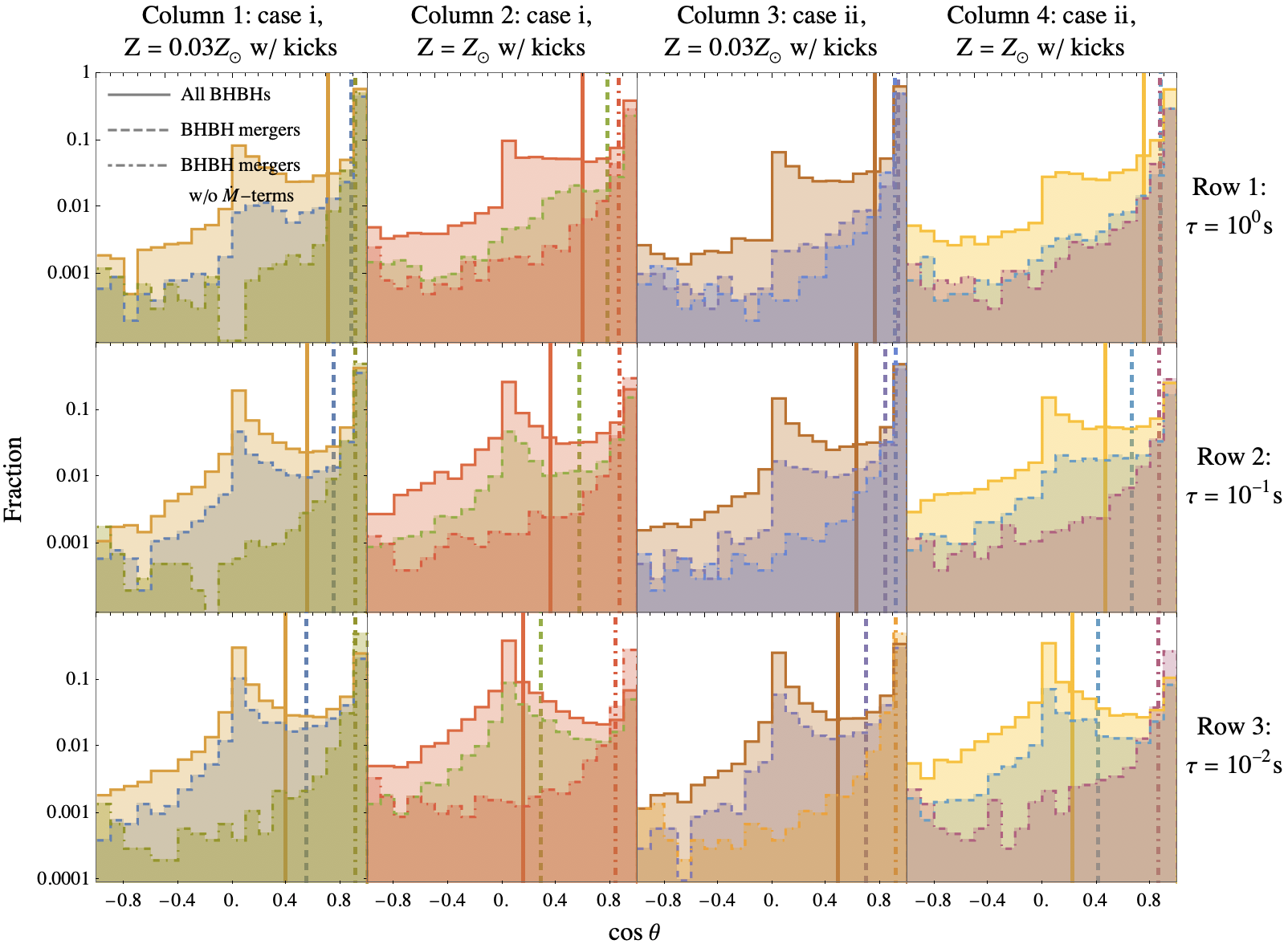}
\caption{Same as Figure~\ref{fig:histo-without-kicks} but natal kicks are included. Additionally, the dot-dashed histograms show the tilt angle distribution of the merging systems ($t_\text{coal}<10\,\text{Gyr}$; dashed histograms) if the spin directions were  {only} affected by the natal kicks and not by the spin dynamics given by Eq.~\eqref{eq:Num-S1}.}
\label{fig:histo-with-kicks}
\end{figure*}

In Figure~\ref{fig:histo-with-kicks}, we take the latter effect into account by implementing the tilts of $\bm{\hat{h}}$ due to the natal kicks at the first and second supernovae. That is, for each supernova we adopt the widely held assumption that the magnitude of the natal kick velocity $v_\text{kick}$ of the black hole follows the one observed for neutron stars scaled down by some fallback-fraction $f_\text{fb}$ \citep{2020A&A...639A..41B},
\begin{equation}\label{eq:kick-vel}
    v_\text{kick}=v_\text{kick,NS}(1-f_\text{fb}),
\end{equation}
where $0\leq f_\text{fb}\leq1$ and $v_\text{kick,NS}$ is drawn from a Maxwellian distribution with a velocity dispersion 
$\sigma_\text{kick,NS}=265\,\text{km}\,\text{s}^{-1}$ \citep{2005MNRAS.360..974H}. Assuming the supernova explosion occurs instantaneously, Eq.~\eqref{eq:kick-vel} can be used to derive the angle $\nu$ between the new and old angular momentum before and after the supernova, respectively \citep[see appendix of][]{2002MNRAS.329..897H}. Thus, the new tilt angle $\theta_\text{new}$ of the primary's spin can be computed as
\begin{equation}
    \cos\theta_\text{new}=\cos\phi\sin\theta_\text{old}\sin\nu+\cos\theta_\text{old}\cos\nu,
\end{equation}
where $\theta_\text{old}$ describes the tilt angle before the supernova and $\phi$ is an angle drawn from a uniform distribution between $0$ and $2\pi$ reflecting an isotropic kick distribution \citep[e.g.,][]{2013PhRvD..87j4028G,2016ApJ...832L...2R}. For producing Figure~\ref{fig:histo-with-kicks}, this method has been successively applied for each supernova. For the merging BBHs the effect of kicks alone can be seen from the dot-dashed histograms for which we skipped the spin dynamics given by Eq.~\eqref{eq:Num-S1}. Instead, we directly applied the kick prescription to the initial tilt angles in the range $[0.9,1.0)$ (see above). The resulting distributions have pronounced peaks at $\cos\theta=1$ with an exponential tail ranging down to $\cos\theta=-1$. The effect of the tails is to broaden the distributions with the whole spin dynamics included (solid and dashed histograms) yielding a small fraction of donor spins that have a $\cos\theta$ below zero.

Furthermore, we investigated whether the distributions presented in Figures~\ref{fig:histo-without-kicks} and \ref{fig:histo-with-kicks} are correlated with the chirp mass $M_\text{chirp}={(M_{1}M_{2})^{3/5}{M_{12}^{-1/5}}}$ which the LIGO-Virgo detectors are most sensitive to. As a result, we find that $\cos\theta$ does not depend on $M_\text{chirp}$. Finally, we also investigated the possibility that at the onset of mass transfer the spin direction is isotropically distributed. That is, we drew the initial value of $\cos\theta_0$ from a uniform distribution in the interval $(-1,1)$. As one would expect, we find that the resulting distributions are shifted towards zero by up to $\sim0.2$.

\begin{figure*}
    \centering
    \subfigure{\includegraphics[scale=0.5]{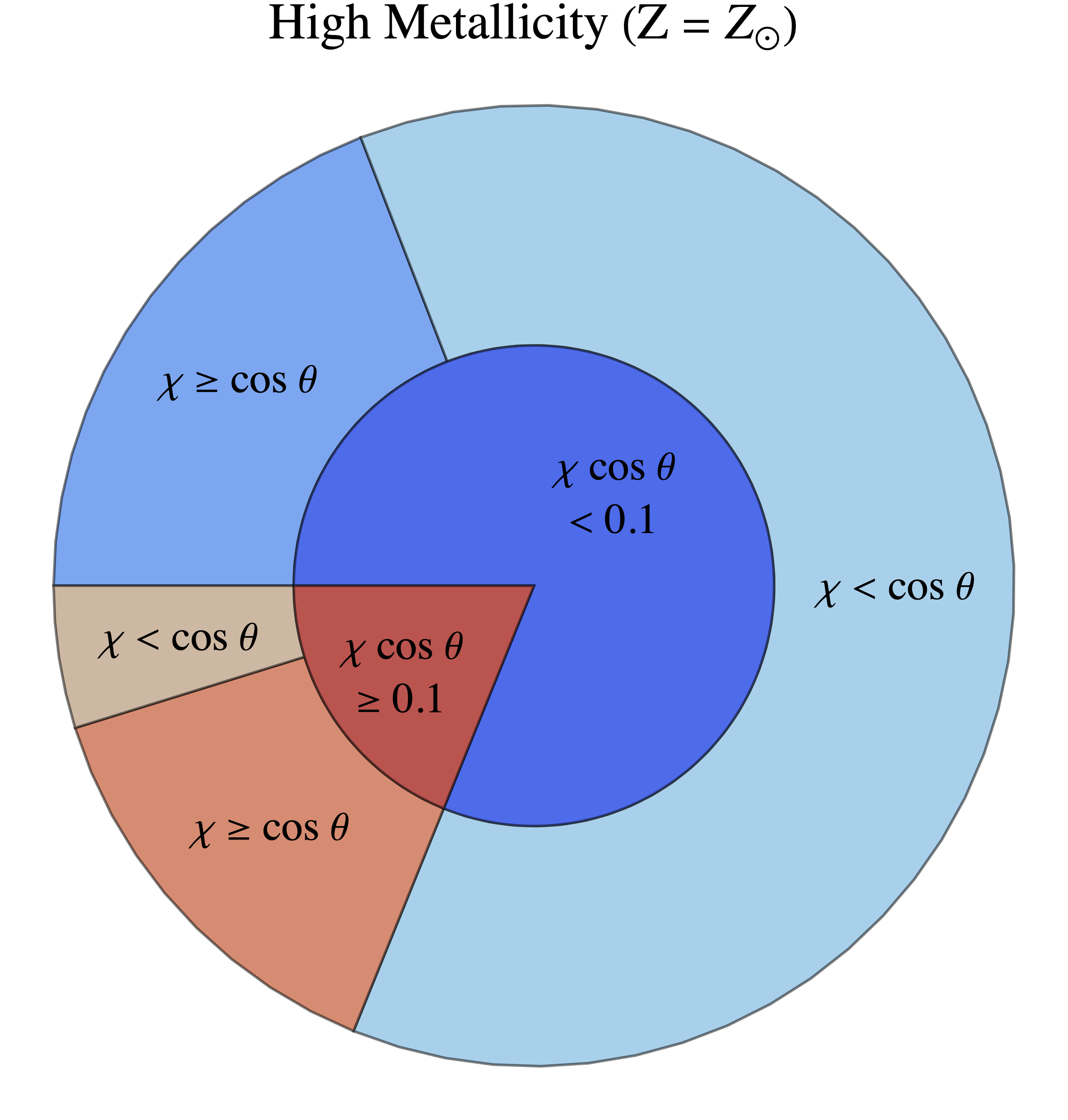}}\quad
    \subfigure{\includegraphics[scale=0.5]{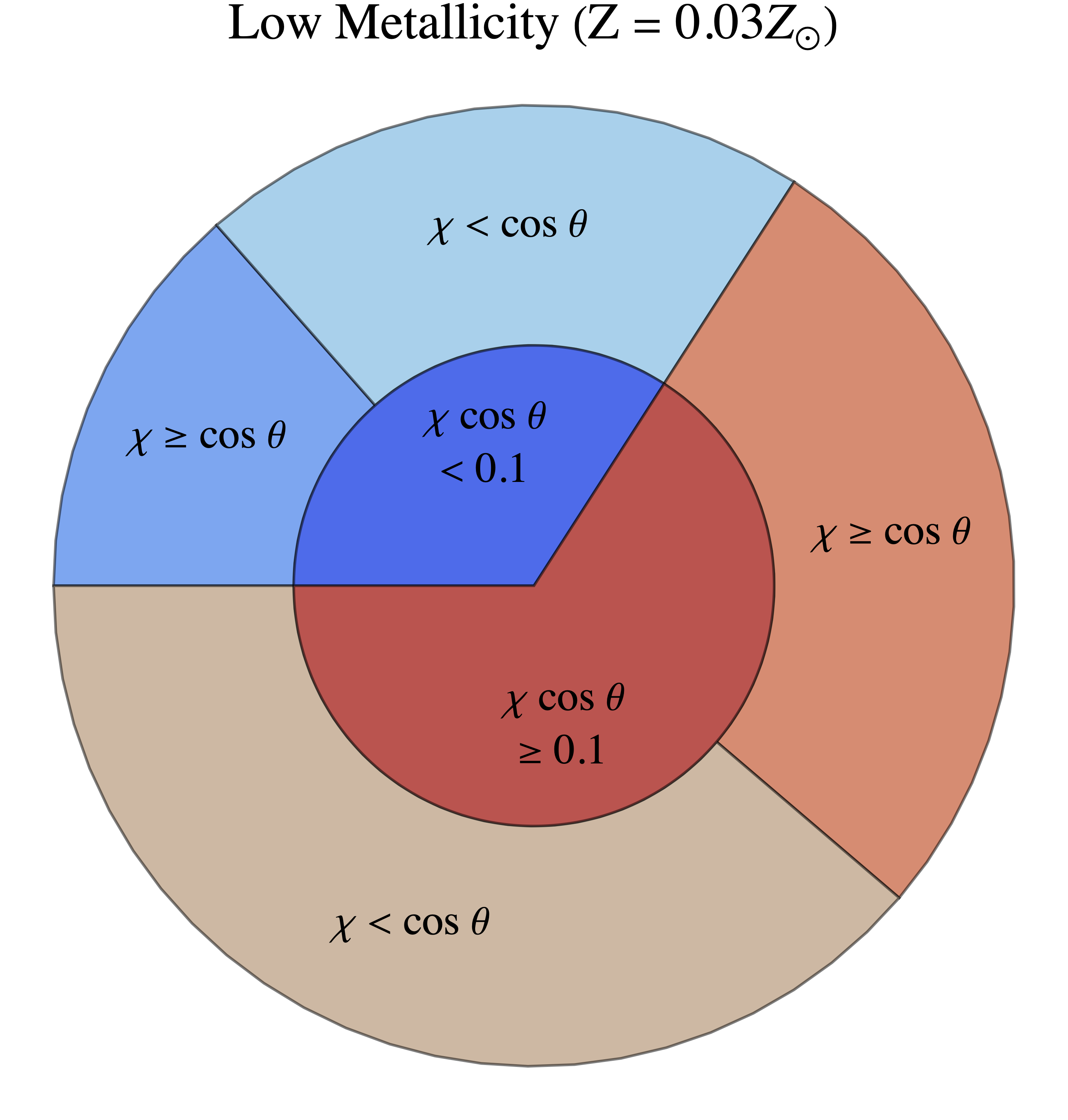}}
    \caption{Spin projections onto the orbital axis of $10^4$ binaries that form {merging} BBHs. The two panels differ from each other by their metallicities. In both cases, mass transfer and tides are modelled with case (i) and a lag time $\tau=10^{-1}\,\text{s}$.}
    \label{fig:pie-chart}
\end{figure*}

Finally, we study the contributions of the first-born black holes to $\chi_\text{eff}$ [cf. Eq.~\eqref{eq:chi-eff}] at its formation. For this purpose, we plot in Figure~\ref{fig:pie-chart} the distribution of the primary spin projection $\chi\cos\theta$ at high ({left}) and low metallicity ({right panel}) for $10^4$ binaries that form merging BBHs. In both cases, mass transfer and tides are modelled with case (i) and a lag time $\tau=10^{-1}\,\text{s}$. The effect of kicks is not taken into account. The inner pies of the panels differentiate the population between primaries whose contribution we consider to be insignificant ($\chi\cos\theta<0.1$) and significant ($\chi\cos\theta\geq0.1$). We stress that this differentiation is somewhat arbitrary, also we do not take the mass-weight into account, but it nevertheless gives a rough estimate of the primaries' spin contribution. We see a clear difference between the two metallicities. Whilst at high metallicity $\chi\cos\theta$ of about 80 per cent is insignificant, this is true for only 30 per cent at low metallicity. This shows that one must not neglect the spin of the first-born black hole if formed at low metallicity. In particular, this result would also hold for stronger tides, i.e. larger values for $\tau$, because in that case $\cos\theta$ would tend to increase towards 1, i.e. align with the orbital axis. For the systems with insignificant contribution we can ask whether this is due to a small magnitude $\chi$ or due to a flip, i.e. a small $\cos\theta$. At both metallicities, the fraction  with $\chi\geq\cos\theta$ is non-negligible. This is true for about 20 per cent and even one half at high and low metallicity, respectively. This suggests that the spin orientation has to be taken into account when studying the spin-contribution of the first-born black hole to $\chi_\text{eff}$ at formation. In Section~\ref{sec:population}, we have shown that the spin magnitude  and its subsequent evolution depend on the spin value at $t=0$. Evidently, the spin orientation becomes irrelevant if we had overestimated the initial spin magnitude, i.e. if $\chi$ were actually smaller than we have assumed. Lastly, we note that during inspiral, the orbit-average of $\chi_\text{eff}$ remains constant at 2PN order whereas the relative contributions of the primary and secondary black hole change due to relativistic effects which become more important as the orbit gets tighter \citep[][]{2001PhRvD..64l4013D,2008PhRvD..78d4021R}. 

\section{Summary}\label{sec:summary}
In this paper, we investigated the vectorial spin evolution of a star that transfers mass to its binary companion. To this end, we modelled the location where the star loses its mass with the intersection point of its surface and the line of separation between the two binary members. Thus, we showed that the mass loss is accompanied with an anisotropic spin-loss that causes the spin magnitude to decrease and its direction to flip onto the orbital plane, i.e. to form a tilt angle of $\theta=\pi/2$ w.r.t. the orbital angular momentum. Generally, this spin dynamics were described by Eq.~\eqref{eq:zeta-1}.
Provided that all parts of the star are sufficiently coupled by efficient angular momentum transport it follows that the solution to this equation also determines the spin direction of all individual parts of the star.

We derived the orbit averaged equations of motion describing the evolution
of the donor spin, assuming either a constant
mass transfer rate per orbit, Eqs.~\eqref{eq:theta-dot}--\eqref{eq:phi-dot}, or a delta-function mass transfer at periapsis, Eqs.~\eqref{eq:theta-dot-delta}--\eqref{eq:phi-dot-delta}. Whilst the former case holds for approximately circular orbits the latter may be a valid model for highly eccentric systems. By considering these two extreme cases we expect that, in reality, the mass transfer rate on moderately eccentric orbits lies somewhere in the intermediate range. 
As a key result of both cases, we found that for total relative mass losses of about $\mathcal{O}(0.1)$ the spin flip is highly efficient unless it starts perfectly or nearly (anti-)aligned with the orbital angular momentum (see Section~\ref{sec:spin-evolution} and Figures~\ref{fig:donor-evolution-constant-mass-rate} and \ref{fig:donor-flip-delta-mass-rate}). Meanwhile, the relative loss in spin magnitude is about several orders of magnitude. As a corollary, we found that both effects are independent of the actual duration of mass transfer but only depend on the total mass that is lost. 

We compared the timescale for spin misalignment due to mass loss to the synchronisation timescale due to tidal torques.  Whether the former effect is faster than tides depends strongly on the stellar separation at the onset of mass transfer, with smaller separations favoring tides.  
We found, however, that the effect of mass loss can   dominate in both main-sequence and giant stars and for a wide range of donor masses. Hence, the commonly adopted assumption that tides will very rapidly erase  any spin-orbit misalignment in mass transferring binaries is not fully  justified.

In reality, a donor might actually have expanded so much that it loses mass trough the outer Lagrangian point too, i.e. through the second ($L_2$) or third ($L_3$) depending on whether it is the less or more massive binary member, respectively. By simulating the response of giant donors to mass loss, \citet[]{2015MNRAS.449.4415P} showed that there exists a critical mass ratio below which its mass is transferred solely and stably through $L_1$. For example, they found that a $30\msun$ donor undergoes $L_2/L_3$ overflow only if it is about two to three times as massive as its companion. If we replace $\bm{R}$ in Eq.~\eqref{eq:zeta-1} by some vector pointing to $L_2/L_3$, i.e. along $-\bm{\hat{d}}$, we see that $\bm{\dot{S}}$ remains invariant up to some positive factor accounting for the larger expansion of the star. Hence, we expect $L_2/L_3$ overflow to promote a spin flip as well. However, $L_2/L_3$ overflow of the donor is typically very short so that the stream of matter is negligible compared to that through $L_1$ \citep[][]{2017MNRAS.465.2092P,2015MNRAS.449.4415P}.

As a potential application, we investigated the spin evolution of isolated stellar binaries which  form a BBH (see Section \ref{sec:population}). A fraction of these BBHs lead to a merger detectable by LIGO-Virgo though their emission of gravitational waves. These binaries have to move on a close orbit making them prone to undergo a phase of stable mass transfer in which the stellar progenitor of the first-born black hole typically loses up to about half of its mass to its companion. By means of a population synthesis, we followed the spin evolution of the primary stars in a large number of BBH forming binaries. To this end, we let our spin dynamics [Eq.~\eqref{eq:zeta-1}] compete with external torques emerging from the quadrupolar distortion of the donor and the tidal interaction between the binary members whose strength was parametrised by the constant lag times $\tau=10^0$, $10^{-1}$, and $10^{-2}\,\text{s}$ [Eq.~\eqref{eq:tvis}]. The stellar physics was simulated at low ($Z=0.03\zsun$) as well as high metallicity ($Z=\zsun$). We found that the resulting tilt angle distribution is strongly bimodal with most spins ending up either aligned with or perpendicular to the orbital angular momentum. The ratio of aligned and flipped systems, however, depends on the metallicity, the tidal lag time, and whether the BBHs merge within $10\,\text{Gyr}$ or not. For instance, going from the long to the short lag time we found that for mergers with low (high) metallicity the $\cos\theta$-distributions' means decrease from $\sim0.9$ ($0.8$ to $0.9$) to $\sim0.5$ to $0.7$ ($\sim0.3$ to $0.5$), i.e. from fair alignment to a mature flip. The values were even smaller by $\sim0.1-0.2$ when we considered all systems that form BBHs. Finally, we have argued that natal kicks are the only effect that could significantly change again the tilt angle between the end of the mass transfer phase and the BBH formation. Whilst the bulk of primary spins remains largely unaffected, natal kicks introduce an exponentially suppressed fraction of primaries with tilt angles towards anti-alignment ($-1<\cos\theta<0$). Overall, we found that at formation the first-born black hole's contribution to $\chi_\text{eff}$ [see Eq.~\eqref{eq:chi-eff}] of a BBH which will merge through the channel considered is typically negligible not only due to its depleted spin magnitude but also due to its misalignment $\cos\theta\sim\mathcal{O}(0.1)$ w.r.t. the orbital axis.

\section*{Acknowledgements}
We thank the anonymous referee and the internal reviewer of the LIGO-Virgo collaboration, Vicky Kalogera, for their suggestions, which helped us to improve this work. We wish to thank Fani Dosopoulou for insightful comments and discussions about the vast literature work on this subject. We acknowledge Ilya Mandel, Davide Gerosa, Michela Mapelli, and Christopher Berry for helpful advice and input and acknowledge the support of the Supercomputing Wales project, which is part-funded by the European Regional Development Fund (ERDF) via Welsh Government. For the numerical simulations we made use of {\fontfamily{qcr}\selectfont GNU Parallel} \citep{tange_ole_2018_1146014}. FA acknowledges support from a Rutherford fellowship (ST/P00492X/1) from the Science and Technology Facilities Council.

\section*{Data Availability}
The data underlying this article will be shared on reasonable request to the authors.

\appendix

\section{Tides and Rotation}\label{sec:apppendix-A}
The perturbations of the equations of motion~\eqref{eq:diff-S}--\eqref{eq:diff-h} can be expressed in terms of five functions $X$, $Y$, $Z$, $V$, and $W$ \citep{2001ApJ...562.1012E,2007ApJ...669.1298F}:

\begin{align}
    X=&-\frac{M_2k_{\rm A}R^5}{\mu na^5}\frac{(\bm{\omega}\cdot\bm{\hat{h}})(\bm{\omega}\cdot\bm{\hat{e}})}{j^4}\nonumber\\
    &-\frac{\bm{\omega}\cdot\bm{\hat{v}}}{2nt_\text{F}}\frac{1+(9/2)e^2+(5/8)e^4}{j^{10}},\\
    Y=&-\frac{M_2k_{\rm A}R^5}{\mu na^5}\frac{(\bm{\omega}\cdot\bm{\hat{h}})(\bm{\omega}\cdot\bm{\hat{v}})}{j^4}\nonumber\\
    &+\frac{\bm{\omega}\cdot\bm{\hat{v}}}{2nt_\text{F}}\frac{1+(3/2)e^2+(1/8)e^4}{j^{10}},\\
    Z=&\frac{M_2k_{\rm A}R^5}{\mu na^5}\Bigg[\frac{2(\bm{\omega}\cdot\bm{\hat{h}})^2-(\bm{\omega}\cdot\bm{\hat{e}})^2-(\bm{\omega}\cdot\bm{\hat{v}})^2}{2j^4}\\
    &+\frac{15GM_2}{a^3}\frac{1+(3/2)e^2+(1/8)e^4}{j^{10}}\Bigg],\\
    V=&\frac{9}{t_\text{F}}\Bigg[\frac{1+(15/4)e^2+(15/8)e^4+(5/64)e^6}{j^{13}}\nonumber\\
    &-\frac{11\bm{\omega}_1\cdot\bm{\hat{h}}}{18n}\frac{1+(3/2)e^2+(1/8)e^4}{j^{10}}\Bigg],\\
    W=&\frac{1}{t_\text{F}}\Bigg[\frac{1+(15/2)e^2+(45/8)e^4+(5/16)e^6}{j^{13}}\nonumber\\
    &-\frac{\bm{\omega}_1\cdot\bm{\hat{h}}}{n}\frac{1+3e^2+(3/8)e^4}{j^{10}}\Bigg],
\end{align}
where the tidal friction timescale 
$t_{\rm F}$ depends on the dissipative mechanism at work as described in Section~\ref{sec:TR}.

\bibliography{apssamp}

\end{document}